\documentclass{article}
\usepackage[utf8]{inputenc}
\usepackage{amsmath, amssymb}
\usepackage[english]{babel}
\usepackage{hyperref}
\usepackage{subfig}
\usepackage{graphicx}
\usepackage{comment}
\usepackage{xcolor}
\usepackage{aas_macros}
\usepackage{authblk}
\usepackage[authoryear]{natbib}

\title{The Background Environment - Solar and Stellar Activity}
\author[1,2]{Stefano Bellotti}
\author[3]{Julien Morin}
\affil[1]{Leiden Observatory, Leiden University, PO Box 9513, 2300 RA Leiden, The Netherlands}
\affil[2]{Institut de Recherche en Astrophysique et Plan\'etologie, Universit\'e de Toulouse, CNRS, IRAP/UMR 5277, 14 avenue Edouard Belin, F-31400, Toulouse, France}
\affil[3]{Laboratoire Univers et Particules de Montpellier,Universit\'e de Montpellier, CNRS, F-34095, Montpellier, France}
\date{}
\begin{document}

\maketitle
%\tableofcontents

\begin{abstract}
This chapter provides an overview of the magnetic activity of the Sun and stars, discussing its underlying physical origin, manifestations, and fundamental role in exoplanet studies. It begins with a summary of the Sun’s magnetic activity from the surface towards the outer atmospheric layers, highlighting features such as sunspots, faculae, chromospheric structures, and their temporal modulation known as the activity cycle. These phenomena are sustained throughout the lifetime of the Sun by the magnetic dynamo, which is driven by differential rotation and convective flows. Furthermore, extending these concepts to other stars, the chapter examines the diagnostics that are typically employed to track and quantify the magnetic activity level of stars, and it reviews spectropolarimetry, an observational technique with which to characterise stellar magnetic fields. We finally outline results from both observations and theoretical modelling of stellar activity across distinct spectral types, and we describe the variety of methods used to search for stellar activity cycles, underscoring the multi-wavelength nature of this field of research. 
\end{abstract}

\section{Introduction}\label{sec:intro}

The most prolific techniques to detect exoplanets are indirect, which means that they rely on a planet's influence on starlight over time, rather than measuring planetary observables \citep{Perryman2018}. The natural implication is that the better we know stars, the better we can study exoplanets. Stellar activity is of paramount importance in this context and it generally refers to the ensemble of dynamo-generated magnetic phenomena occurring on the stellar surface and outward in the atmosphere. Distinguishing the signal of a planet from the host star is key to a successful detection and characterisation, and relies on accurate knowledge of the stellar surface flux inhomogeneities. 

The Sun’s magnetic activity is a dynamic and complex process that governs many of the phenomena observed on its surface and in its outer layers. At the heart of this activity is the solar dynamo, a mechanism driven by the interplay between differential rotation and convective motions in the Sun’s interior \citep{Charbonneau2020}. This process generates the Sun’s magnetic field, which undergoes cyclic variations, most notably the 11-year sunspot cycle, during which the number and distribution of sunspots fluctuate \citep{Hathaway2015}. Understanding solar activity is critical for refining our knowledge of the stellar magnetic processes. The solar dynamo, while well-studied, remains an area of active research, with numerical simulations striving to reproduce observed magnetic cycles \citep{Charbonneau2020,Kapyla2023}. Beyond the photosphere, the influence of the solar magnetic field extends to the corona, where it drives high-energy processes such as solar flares and coronal mass ejections \citep[see e.g.][]{Webb2012}. These events can significantly impact the solar wind and space weather, affecting planetary environments, including the Earth’s magnetosphere \citep{Georgieva2023}.

Like the Sun, many stars exhibit primary manifestations of magnetic activity like starspots, high-energy events, and magnetic cycles \citep{Strassmeier2009,Kowalski2024,Jeffers2023}. Stellar activity is a fundamental property across various spectral types, and its features vary with stellar characteristics such as mass, rotation, and internal structure \citep{Brun2017}. At the same time, stellar magnetic fields are tightly linked to the high-energy radiation and stellar winds which, in turn, regulate the space environment in which exoplanets orbit. This implies that stellar magnetic activity has a direct influence on the evolution of exoplanet atmospheres over time as well as their habitability conditions \citep[e.g.][]{Lammer2003,Segura2010,Vidotto2014,Tilley2019}. Therefore, studying activity for stars across different spectral types allows one to build a comprehensive framework for stellar magnetism and its role in the evolution of planetary systems.

\section{Solar magnetic activity}\label{sec:solar_activity}

\begin{figure}[!t]
\centering
\includegraphics[width=0.325\textwidth]{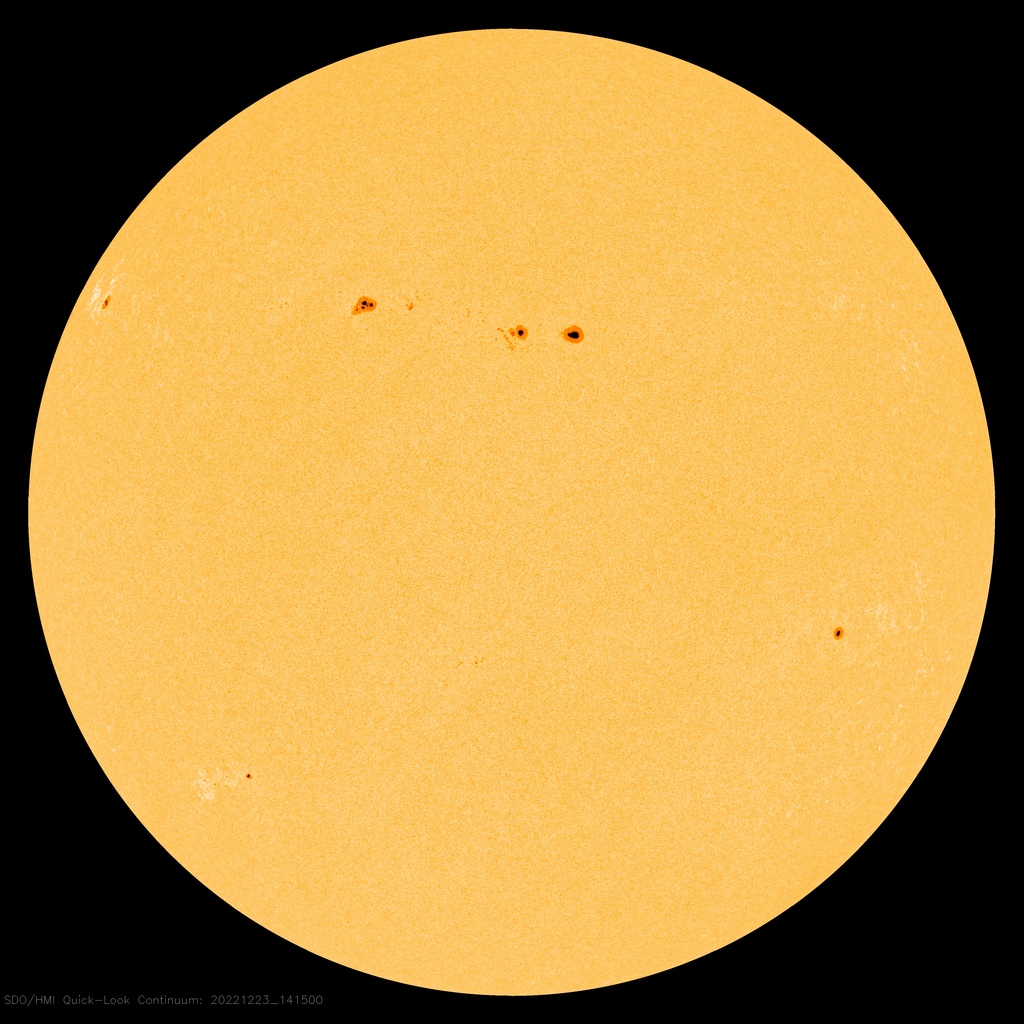}
\includegraphics[width=0.325\textwidth]{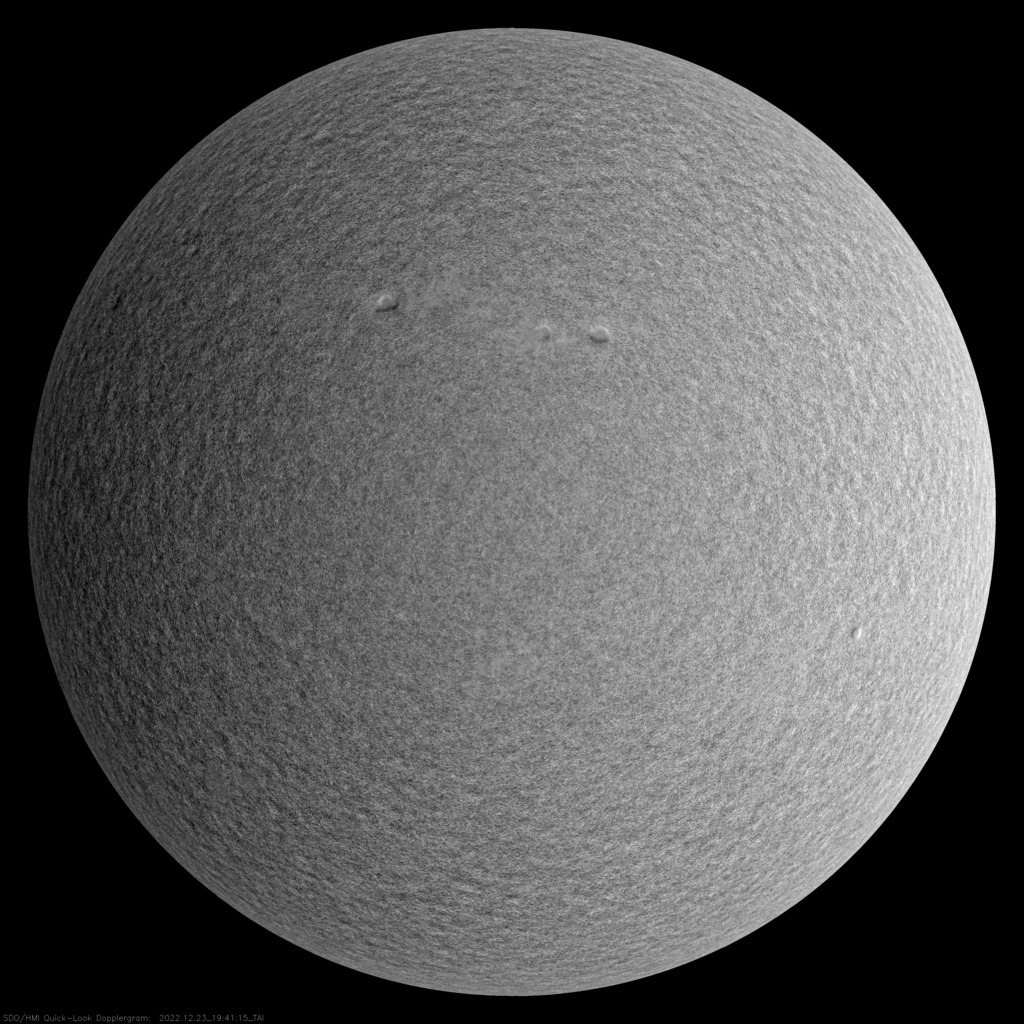}
\includegraphics[width=0.325\textwidth]{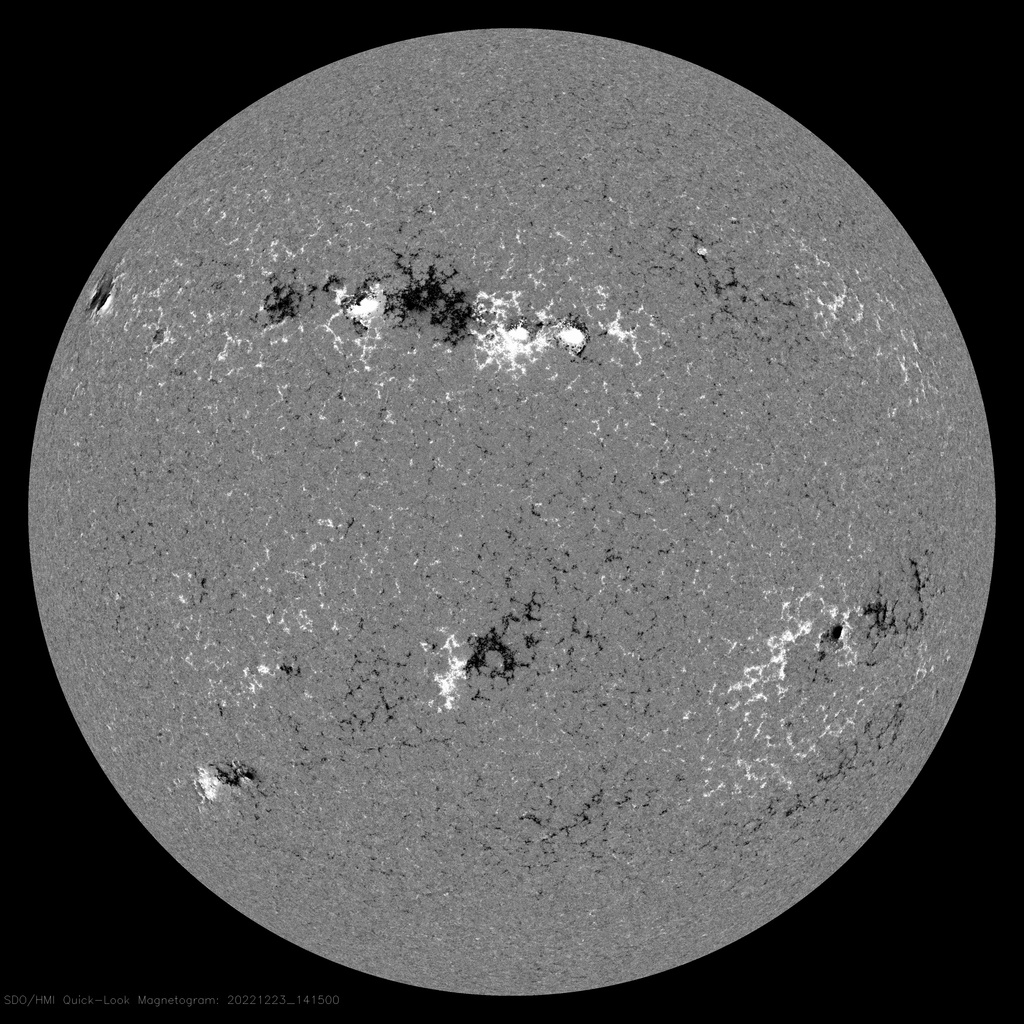}
\caption{Images of the Sun in December 2022 for the flattened brightness (left), line of sight velocity (middle), and magnetic field (right) acquired with the Solar Dynamic Orbiter. Examples of visual stellar activity manifestations are photospheric inhomogeneities, like faculae and spots which are hotter and cooler than the rest of the surface, respectively. These are connected to the underlying magnetic field. Credit: NASA/SDO and the AIA, EVE, and HMI science teams.}
\label{fig:sun}
\end{figure}

Signatures of solar and stellar activity at optical and near-infrared wavelengths originate from the photosphere and chromosphere (see Fig.~\ref{fig:sun}). The photosphere is the bottom layer of the stellar atmosphere and on the Sun it features an effective temperature of 5772\,K \citep{Mamajek2015}. Above the photosphere lies the chromosphere, which exhibits a temperature minimum at a height of 500\,km from $\tau_{500}=1$ (where $\tau_{500}$ is the radial optical depth in the continuum at 500\,nm) and a steep temperature rise to a few $10^4$\,K around 2000\,km \citep[see Fig.~3 in][]{Hall2008}. The chromosphere is an heterogeneous layer with a complicated magnetic topology, and with several evanescent features that vary on time scales on the order of minutes. 

Thanks to the high spatial and temporal resolution available, solar observations have revealed a plethora of photospheric heterogeneities such as seething granules (which are not magnetic in nature, but due to convection), dark sunspots, and bright faculae \citep[see][for a review on the solar active regions]{Schrijver2000}. Studies revealed that these active regions originate from the emergence of magnetic flux tubes on the solar surface \citep{Schuessler2002}, whose magnetic field is intense enough (on the order of kilo-Gauss) to alter convective motions and inhibit the energy transport \citep{Schrijver2000,Solanki2003,Strassmeier2009}. Sunspots temperatures range between 4000~K and 5000~K and their magnetic field strengths between 1~kG and 4~kG \citep{Solanki2003}. They also tend to cluster in groups, called `nests', which exhibit complex structures and dynamics \citep[see e.g.][]{Pojoga2002}. The emergence of narrow magnetic flux tubes gives origin also to bright photospheric and chromospheric features known as faculae and plages, respectively, which are best observed at the solar limb \citep{Schrijver2000}. They can be 300 to 500\,K hotter than the photosphere \citep{Topka1997} and, although they can appear separately from spots and clump to form a larger feature, they tend to surround spots. Depending on whether they are isolated or clumped in groups, their duration spans between hours or days \citep{Hirayama1978,Solanki1993}.

\begin{figure}[!t]
\centering
\includegraphics[width=\textwidth]{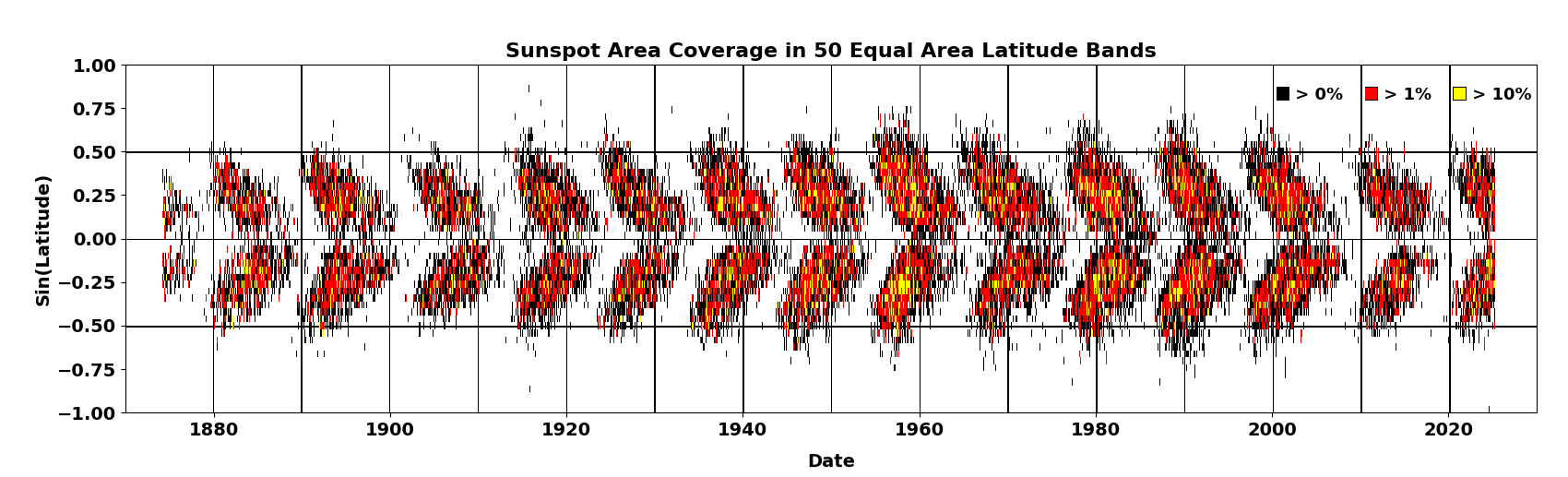}
\caption{The butterfly diagram illustrates the distribution of sunspots and magnetic flux on the Sun and their change over time. The figure shows the sunspots area coverage as a function of latitude, for which it is possible to see the Sp\"orer law, that is the equatorward drift of the sunspot emergence through a magnetic cycle. Credit: Dr. D. Hathaway \href{http://solarcyclescience.com/solarcycle.html}{http://solarcyclescience.com/solarcycle.html}.}
\label{fig:butterfly}
\end{figure}

In 1843, S. H. Schwabe noticed that the distribution of sunspots varies periodically over a time scale of 11\,yr \citep{Schwabe1844}, a phenomenon today known as {\it sunspot cycle} (or Schwabe cycle). During the maximum of the cycle, the sunspot number is highest and larger sunspots appear, while the opposite occurs at cycle minimum. Typically, the monthly-averaged number of sunspots is around 0-20 at minimum and around 150-200 during maximum, but the exact number varies across different cycles \citep{Hathaway2015}. At the start of the cycle, sunspots appear at around $35^{\circ}$ latitude on either side of the solar equator and, over time, they emerge progressively closer to the equator. Such tendency to form at lower latitudes is called the Sp\"orer law, and was initially captured with observations by \citet{Maunder1904} with the well-known butterfly diagram (see Fig.~\ref{fig:butterfly}). The periodicity of the solar cycle is not exact, since variations in both its amplitude and length have been observed, together with epochs of significantly lower activity known as Maunder minimum \citep[][]{Maunder1894}.

\begin{figure}[!t]
\centering
\includegraphics[width=0.49\textwidth]{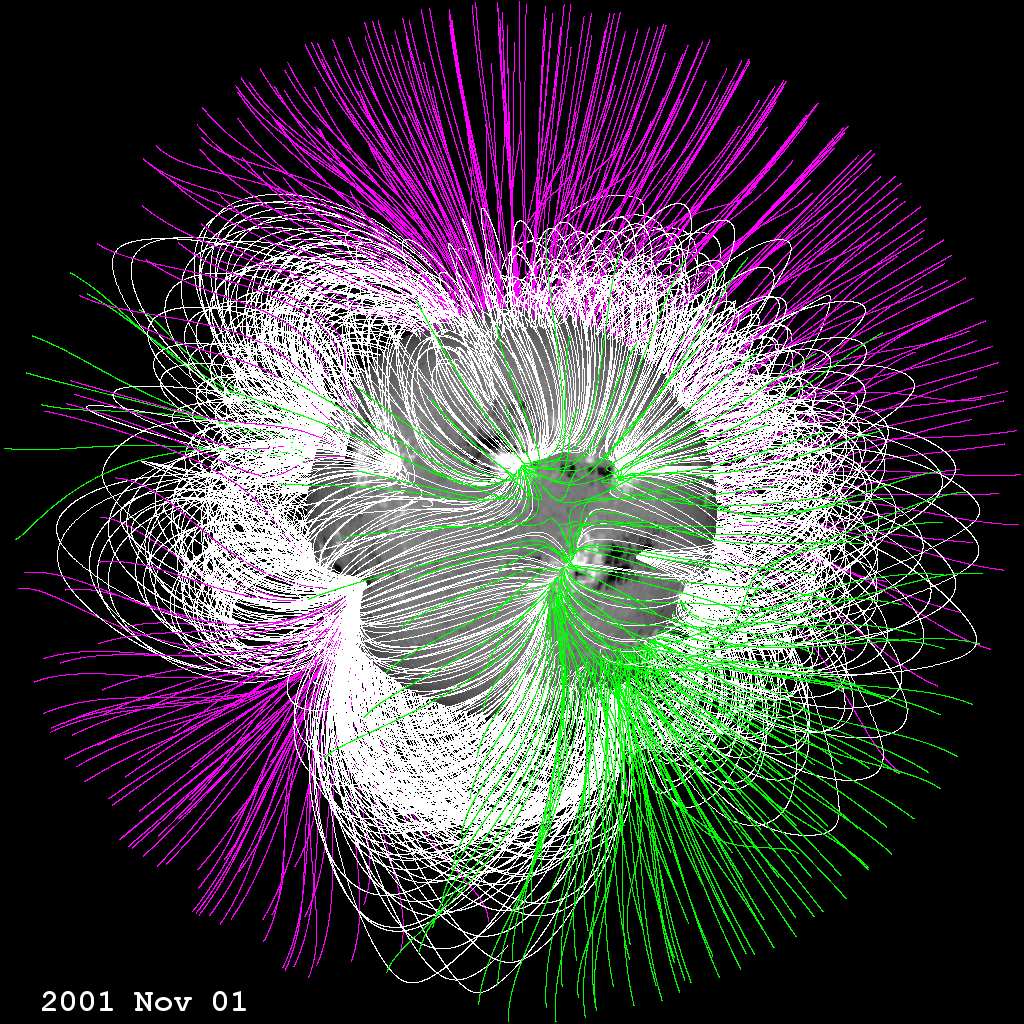}
\includegraphics[width=0.49\textwidth]{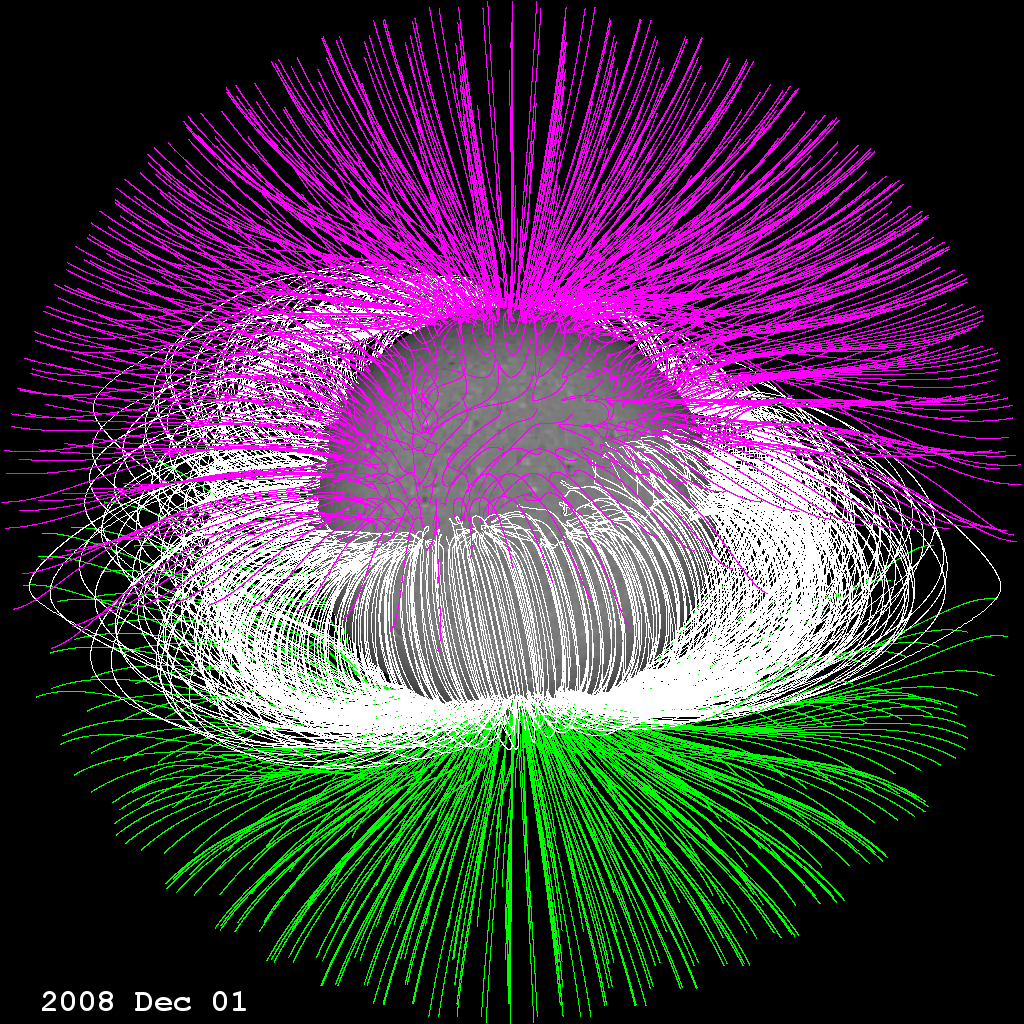}
\caption{Potential Field Source Surface (PFSS) model of the Sun during cycle maximum and minimum. The images are built using the SolarSoft package from solar magnetograms collected by means of SOHO/MDI and SDO/HMI instruments, and the extrapolation is performed from the photosphere out to about 2.5 solar radii, which is where the source surface is located. The left image (November 2001) illustrates the magnetic field of the Sun at activity maximum (during cycle 23) and the right image (December 2008) illustrates the magnetic field at the activity minimum (and the start of cycle 24). Purple and green colours indicate open field lines of negative and positive polarity, while the white colour indicates closed field lines. The large-scale magnetic field has a simple, mostly dipolar configuration at activity minimum, and a complex one at activity maximum. Credit: NASA's Goddard Space Flight Center Scientific Visualization Studio/Bridgman and Duberstein.}
\label{fig:sun_field}
\end{figure}

These findings were later complemented by the discovery of the magnetic nature of sunspots by \citet{Hale1908} and the polarity reversal laws \citep{Hale1919}, revealing the {\it magnetic activity cycle} of 22~yr (also known as the Hale cycle). For 11\,yr the majority of the sunspots on the northern hemisphere have the same polarity, whereas most of the southern hemisphere sunspots have the opposite polarity. Such configuration switches during the next cycle around maximum, making the magnetic cycle composed by two consecutive sunspot cycles \citep[see the reviews of][]{Hathaway2010,Hathaway2015}. On a global scale, the magnetic field of the Sun is predominantly dipolar during cycle minimum, while during maximum it becomes more complex, since more energy is stored in quadrupolar and octupolar modes (see Fig.~\ref{fig:sun_field}). In parallel, the amount of magnetic energy in the poloidal and toroidal large-scale field components varies during a cycle, and the obliquity of the poloidal-dipolar component oscillates between axisymmetric and non-axisymmetric configurations \citep{Babcock1955,Sanderson2003,DeRosa2012,Vidotto2018,Finley2023}.

Above the chromosphere, there is a narrow (about 100\,km) layer called the transition region where the temperature increases from $5\cdot10^4$\,K to $\sim10^6$\,K. The outer region is the corona, which contains hot ($\sim3\cdot10^6$K) and ionised plasma extending out into space in the form of a supersonic outflow known as the solar wind. The fact that the corona is hotter than the photosphere seems to challenge thermodynamics, since one expects the temperature to decrease away from the heat source (the Sun's core in this case), an aspect known as the `coronal heating issue'. Although a definitive solution has not yet been found, the underlying heating process may involve a combination of different mechanisms such as magnetohydrodynamic waves dissipation, small-scale magnetic reconnections, and turbulence \citep[see][for a recent review]{Cranmer2019}.

The corona is recognisable to the naked eye during total solar eclipses as a white halo surrounding the solar disk as consequence of Thompson scattering of free electrons. This layer of the solar atmosphere is observed mostly at radio wavelengths and in X-rays \citep{White1999}, which are generated from collisionally-excited ions in magnetically confined plasma, heated by magnetic fields to temperatures of several million Kelvin \citep{Vaiana1981}. The solar spectrum in the X-ray range ([1, 100]~{\AA}) is dominated by coronal emission, while in the extreme ultraviolet (EUV, [100, 912]~{\AA}) the emission is originated both in the corona and in the transition region \citep{Phillips2008}. Observations at these wavelengths provide information on density and structural complexity of the solar corona. Studies showed that the X-ray and EUV emission varies by one order of magnitude in correlation to the magnetic cycle \citep{Peres2000,Woods2005}, and it is therefore driven by the solar magnetic dynamo (see the Sect.~\ref{sec:solar_dynamo}).

Studying the high-energy radiation emitted in solar and stellar coronae, specifically X-rays and EUV, is crucial to determine the state of a planet’s atmosphere. Such high-energy radiation reaches different heights of the planetary atmosphere (upper layers for EUV and deeper for X-rays) causing ionisation, dissociation, and heating phenomena, which enhance and drive atmospheric loss processes \citep[e.g.][]{CecchiPestellini2006,Penz2008b}. Overall, studying stellar high-energy radiation provides a window to understand and different evolution pathways after formation \citep[e.g.][]{Lammer2003,LopezFortney2013,Owen2012,France2013,Owen2017}. This will be discussed in more detail in Chapter 14.

\subsection{The solar dynamo theory}\label{sec:solar_dynamo}

The Sun has a partly-convective interior, meaning that it has a radiative core surrounded by a convective envelope. The energy produced in the core by hydrogen thermonuclear fusion is initially transported outward, through the densest and hottest regions, by radiation. At larger radii, energy is primarily transported by convection: rising cells of hot plasma transport energy to the photosphere, cool down and sink back in a cyclical fashion. The interface between the radiative and convective zone is called the tachocline, a region of strong shear revealed by helioseismic studies \citep{Spiegel1992} and the location where magnetic flux tubes are thought to be stored and amplified until they eventually rise to the surface and generate spots. From a theoretical standpoint, the cyclic regeneration of the solar magnetic field is explained by dynamo theory, namely the active conversion of kinetic energy associated to hydrodynamical motions of plasma in stellar interiors into magnetic energy, mainly via coupling of rotation and convection. For a recent review on dynamo models, see \citet{Charbonneau2020}.

The magnetohydrodynamical induction equation describes the evolution for the magnetic field ${\bf B}$ and can be derived from the Maxwell–Faraday and Maxwell–Amp\`ere equations together with the generalized Ohm’s law. Indicating with ${\bf u}$ the fluid motion vector and with $\eta$ the magnetic diffusivity, the equation reads
\begin{equation}\label{eq:induction}
    \frac{\partial {\bf B}}{\partial t}=\nabla\times({\bf u}\times{\bf B}-\eta\nabla\times{\bf B})=\nabla\times({\bf u}\times{\bf B})+\eta\Delta{\bf B},
\end{equation}
where the first term on the right hand side represents induction and the second term indicates dissipation. The relative importance of the two terms is encapsulated in the magnetic Reynolds number ${\rm Rm}=uL/\eta$, that is the ratio between the induction and diffusion terms, with $u$, $\eta$, and $L$ the characteristic numerical values for flow velocity, magnetic diffusivity, and length scale over which a significant variation of ${\bf B}$ occurs. We note that in practice dynamo simulations require an initial magnetic field seed \citep[e.g.][]{Khomenko2017}, without which the magnetic induction equation would predict zero indefinitely. Two additional equations have to be considered to complete the dynamo problem: the Maxwell's equation $\nabla\cdot{\bf B}=0$ for the magnetic field, and the Navier-Stokes equations for the evolution of the fluid motion vector ${\bf u}$ \citep[see][]{Charbonneau2020}. The dynamo problem involves generating a flow motion ${\bf u}$ that is dynamically consistent and possesses inductive properties to sustain the magnetic field ${\bf B}$ against Ohmic dissipation. For the Sun, this translates into finding the circumstances under which the flow can reproduce the solar magnetic cycle. 

\begin{figure}[!t]
    \centering
    \includegraphics[width=\textwidth]{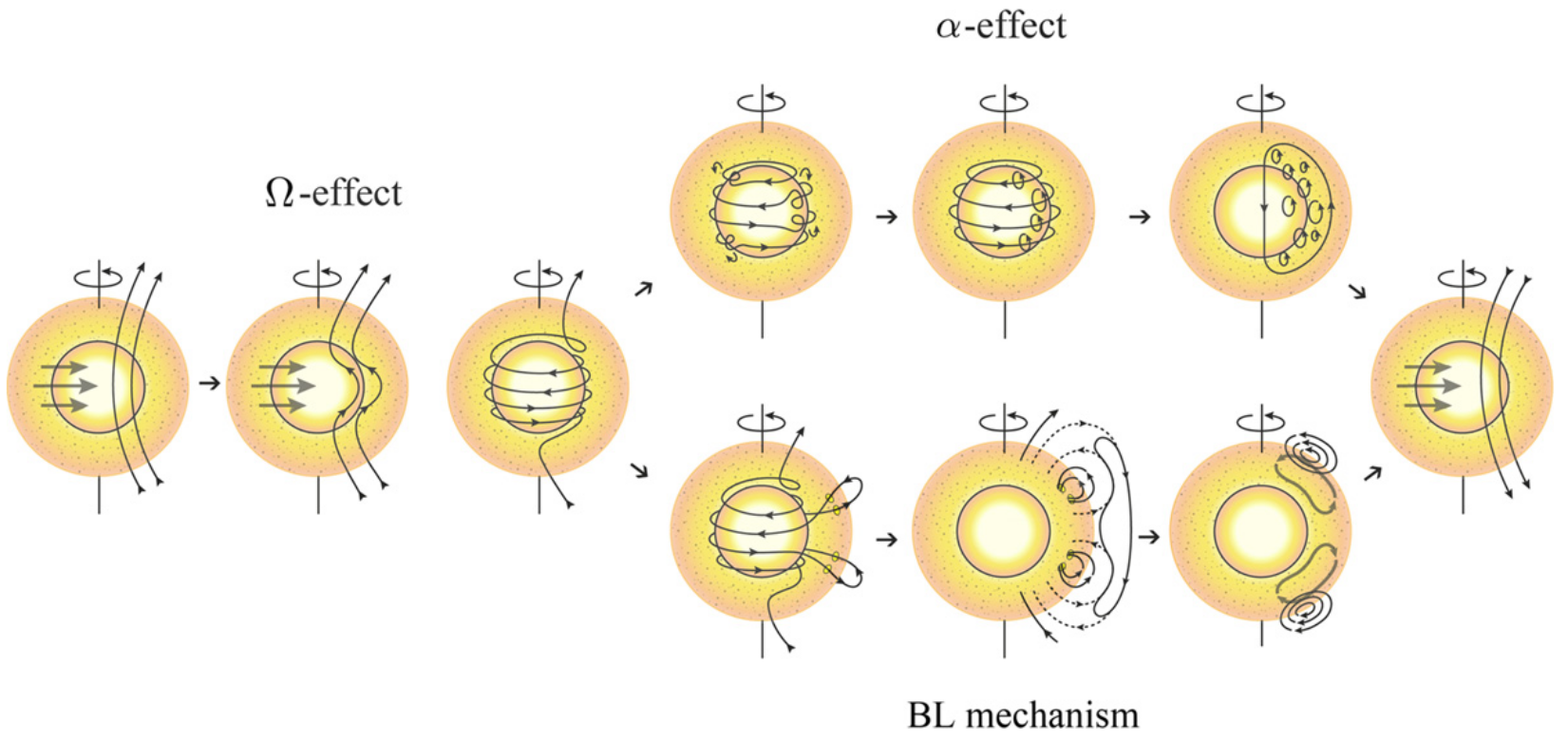}
    \caption{Mechanisms at the base for the solar dynamo model. The $\Omega$-effect transforms a poloidal field into a toroidal field via differential rotation. Restoring a poloidal field from a toroidal field is accomplished in two ways: the $\alpha$-effect sees cyclonic turbulence twisting toroidal field lines into small-scale poloidal fields which, on average, constitute a global poloidal field. The Babcock-Leighton (BL) mechanism sees the formation of bipolar regions on the surface; those close to equator diffuse and reconnect between hemispheres, whereas the remaining ones are transported towards the pole by meridional circulation and produce a large-scale poloidal field. Image credit: \citet{Sanchez2014}.}
    \label{fig:alpha_omega}%
\end{figure}

One scenario is given by the $\alpha\Omega$ dynamo \citep{Parker1955, Charbonneau2010}, that is the interplay between differential rotation ($\Omega$-effect) and cyclonic turbulence ($\alpha$-effect) at the level of the tachocline. The $\Omega$-effect was already proposed by \citet{Larmor1919}, but the issue of poloidal field regeneration remained suspended until the work of \citet{Parker1955}. Fig.~\ref{fig:alpha_omega} illustrates the $\alpha\Omega$ dynamo: 1) differential rotation stretches poloidal magnetic field lines over latitudes, converting them into a toroidal configuration, and 2) small-scale turbulent plasma motions twist toroidal lines into loops which, on large-scales, result in a toroidal current passing through these loops, ultimately regenerating a poloidal field. An alternative to the $\alpha\Omega$ dynamo is the Babcock-Leighton mechanism (see Fig~\ref{fig:alpha_omega}), which describes the conversion from toroidal to poloidal field via a poleward migration of bipolar magnetic regions \citep{Babcock1961,Leighton1969}. In practice, sunspots emerge in pairs and tilted relative to the East-West direction, with the leading sunspot (with respect to the solar rotation direction) at a lower latitude; this phenomenon is known as Joy's law and it has a latitudinal dependence, being stronger at higher latitudes \citep{Stenflo2012}. The tilt of this bipolar magnetic region translates in a non-zero dipole moment, which gradually diffuses via surface flows and reconnects with other regions. Taking into account several magnetic regions, their advection towards the pole via meridional circulation ultimately results in a net large-scale poloidal field. Such processes can generate and sustain magnetic fields against Ohmic decay over stellar lifetimes \citep{Charbonneau2010}. Mean-field dynamo models based on the $\alpha\Omega$ mechanisms can explain the cyclic magnetic activity, polarity reversals, and the role of differential rotation, but they fail at reproducing the observed equatorward migration of sunspots as captured by the butterfly diagram. Babcock-Leighton models are more observationally grounded and explain Hale's and Joy's laws about the polarity and tilt of sunspots, as well as surface magnetic flux transport and the butterfly diagram. However, BL models still have limitations. For example, they usually produce excessively strong magnetic fields at the poles and their dynamo solutions are not self-excited, meaning that they cannot amplify an arbitrarily weak seed magnetic field \citep[see the reviews of][]{Brun2017,Charbonneau2020}.  

Magnetohydrodynamical (MHD) simulations of the solar dynamo in three-dimensional spherical shells corroborated the central role of the tachocline, especially to achieve global-scale ordering of structures \citep{Browning2006,Lawson2015,Guerrero2016}. Other simulations, focusing on the bulk of the convection zone, showed the formation of enduring magnetic wreaths coexisting with turbulent convection and in the absence of a tachocline \citep{Brown2010,Strugarek2017}, a result that put the role of this thin interface into question. In terms of cycles, more recent MHD simulations have succeeded at reproducing polarity reversals and quasi-regular oscillations \citep[e.g.,][]{Brown2011,Guerrero2016,Strugarek2018}, but our understanding of the cyclic nature of the solar magnetism is sill not complete \citep{Charbonneau2020}. Together with the poloidal-toroidal interplay, an ideal model for solar dynamo should encompass additional aspects that characterise the solar magnetism such as the field strength, the equatorward migration of sunspots, the flip-flop cycles, and the dominant hemispheric magnetic helicity \citep[see also the reviews of][]{Charbonneau2020,Kapyla2023}.

Reproducing the solar magnetic cycle and the dynamo loop, that is the alternated generation of poloidal and toroidal field components to sustain the solar magnetism, is an active field of research \citep{Brun2017,Charbonneau2020}. On one side, this can be seen by the challenge in predicting the solar cycle \citep{Petrovay2020}, and on the other side a number of difficulties in numerical simulations of dynamo models remain, such as reproducing the solar convection, the differential rotation, and the saturation of activity as seen with different proxies (this feature will be described in Sect.~\ref{sec:results_activity}). For instance, in the recent review of \citet{Kapyla2023} about solar and stellar dynamos, the authors describe the challenge of simulating the latitudinal differential rotation of the Sun, because the solutions fall into anti-solar regime (slowly-rotating equator, fast-rotating pole) rather than the observed solar regime (fast-rotating equator, slowly-rotating pole). In this context, further progress towards a comprehensive dynamo theory requires both solar observations of magnetic phenomena, with the benefit of high spatial and temporal resolution, and observations of magnetism and cycles on other stars to allow parametric studies of such phenomena.

\section{Stellar magnetic activity}\label{sec:stellar_activity}

\subsection{Starspots}\label{sec:starspots}

Counting the number of spots on the Sun is a direct measurement of solar activity owing to the available spatial resolution (see Fig.~\ref{fig:sun}). This is not possible on other stars as they appear as point-like sources. In this case, one can track the photometric variability of a star to evaluate its activity level, since the temporal modulation of a light curve is assumed to stem from the presence of inhomogeneities on the stellar surface. Depending on the complexity of the light curve, large-amplitude spot modulations and sharp transits of corotating gas clouds may be evoked to explain the photometric modulation \citep[see][]{Gunther2022}, and in some cases it is possible to model the lightcurve to obtain a starspot map \citep[e.g.][]{Ikuta2023}. In recent work, the use of starspot transit mapping has been proposed for the measurement of small-scale magnetic fields. The idea is applicable for stars that are known to host exoplanets and relies on the detectable variations in the transit light curve caused by a planet eclipsing its host star and occulting dark spots at the same time \citep{Silva2003}. Using empirical relations, one can estimate the magnetic field of spots from their flux deficit, that is the product of the spot area and brightness \citep{Valio2020,Menezes2024}.

Distinct photometric activity indicators have been investigated \citep[e.g.][]{Basri2010,Basri2013}, among which the standard deviation of the light curve $S_{ph}$ \citep{Garcia2010} and its refinement by \citet{Mathur2014} to account for temporal evolution of the activity level. The idea is to split the light curve in sections whose length is defined by the stellar rotation period, compute the standard deviation for each, and analyse the temporal variability. The mean standard deviation $\langle S_{ph}\rangle$ represents the global magnetic activity level. Indeed starspots last between few days and several rotation cycles \citep[e.g.][]{Hussain2002,Namekata2019}, and in proportion to their size and the star's spectral type \citep{Berdyugina2005,Giles2017}. Additionally, the presence of outlying data points may indicate an abrupt brightening of the star due to flares (see Chapter~4 for more details).

Spots are a notorious source of variability also for spectroscopic and spectropolarimetric time series \citep[see the reviews of][]{Berdyugina2005,Strassmeier2009}. The balance in the emitted flux between the approaching stellar hemisphere (blueshifted) and the receding one (redshifted) is in fact broken by the presence of a dark spot on the surface. As the spot crosses the visible disk, it reduces part of the integrated starlight and produces distortions in the profile of spectral lines, which translates into Doppler shifts modulated at the stellar rotation period \citep{Saar1997}. For exoplanet hunters using the radial velocity method, these Doppler shifts prevent reliable detections, as they can hide or mimic the presence of a planet. Moreover, spots inhibit convective blueshift, which is an effect due to the imbalance between the number of photons coming from the hot (bright) material relative to the cold (dark) one undergoing convection on the stellar surface. The combination of these spot-induced phenomena represent an important obstacle to overcome for reliable detection of small planets \citep[see the review of][]{Meunier2021}. 

Monitoring the temporal modulation of the distortions induced on spectral lines is at the base of a tomographic inversion technique called Doppler imaging, with which one can reconstruct a 2D map of the stellar brightness \citep{Deutsch1957,Vogt1983,Rice1989}. This is possible thanks to the correlation between the location of the spectral distortion on the profile and the location of the inhomogeneity on the stellar surface, owing to the Doppler effect. The occurrence of a distortion at a certain time yields information on the longitude of the surface inhomogeneity, while the extent of the distortion across the line profile informs one of the latitude of the spot. The longitudinal resolution of the reconstructed map scales with the stellar rotational velocity $v_\mathrm{eq}\sin i$ \citep[e.g.][]{Hussain2009,Morin2008a}. Values of $v_\mathrm{eq}\sin i \geq15-20$ km\,s$^{-1}$ typically ensure sufficient rotational broadening of the line profile and ultimately the reliability of the reconstructed map. Examples of Doppler imaging maps of fast-rotating, main-sequence stars of different ages can be found in the work of \citet{Donati2000,Morin2008a,Donati2016,Cang2020,Finociety2021}. A list of early reconstructions can also be found in \citet{Strassmeier2009}.

Along with Doppler imaging, there are other techniques one can apply on very active stars. Using spectroscopic observations of molecular lines, it is possible to characterise starspots temperature and filling factor, that is the area coverage of the stellar surface. The results of \citet{ONeal2004} and \citet{ONeal2006} are consistent with Doppler imaging findings. They modelled the absorption bands of TiO taking a combination of spectra of inactive M stars as proxies for the active photosphere and spectra of G and K stars as proxies for the unspotted photosphere. The strength of the TiO bands of each proxy has a different temperature sensitivity, hence one can constrain the starspot temperature, and the absolute strength of the bands correlates with the projected area of the starspots. Each proxy is weighted by a spot filling factor and the continuum surface flux ratio to reproduce spectra of active stars.

Following the first measurement of magnetic fields within starspots for M~dwarfs \citep{Berdyugina2006}, \citet{Afram2019} estimated magnetic fields at distinct atmospheric depths adopting certain chemical species. They established that TiO and FeH form in the upper and lower atmosphere, respectively, whereas atomic lines at intermediate heights. Furthermore, the lower altitude atmosphere of early-M dwarfs was reported to feature simpler magnetic spots than later M~dwarfs, while the upper atmosphere has more complex structures regardless of the spectral type. On the contrary, \citet{Shulyak2017,Shulyak2019} found consistent values of the magnetic field for atoms and molecules, implying that further investigations are needed.

\subsection{Activity diagnostics}\label{sec:measure_activity}

\begin{figure}[!t]
    \centering
    \includegraphics[width=\textwidth]{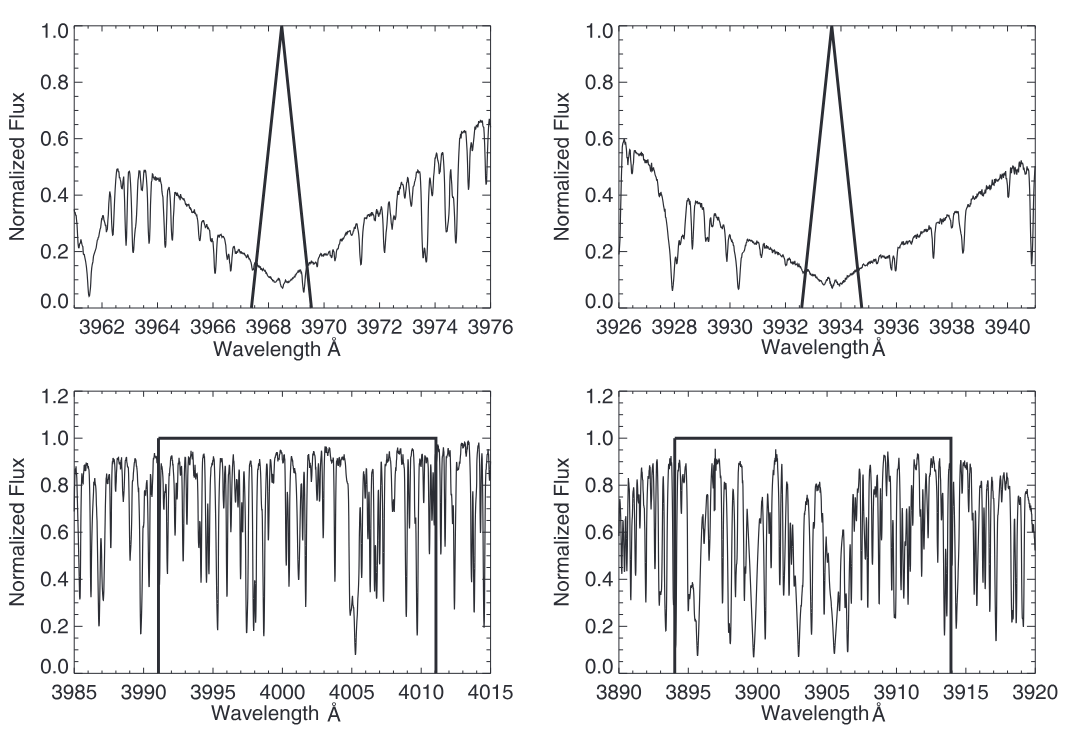}
    \caption{Spectrum around the Ca~\textsc{ii} H\&K lines and spectral windows defining the $S$-index. Top row: the Ca~\textsc{ii} H\&K lines (left and right, respectively) with triangular windows whose full width at half maximum is 1.09~{\AA}. Bottom row: two continuum sections on either side of the H and K lines, identified as V (centred at 4000 {\AA}) and R (centred at 3905 {\AA}), with two 20 {\AA} windows. Image credit: \citet{Isaacson2024}.}
    \label{fig:CaIIHK}%
\end{figure}

Quantifying the activity level of stars requires indicators. The H\&K lines of the doubly ionised calcium at 3968.470\,{\AA} and 3933.661\,{\AA} have been used extensively to quantify the magnetic activity of solar-like stars during the Mt. Wilson project \citep{Wilson1968}, as shown in Fig.~\ref{fig:CaIIHK}. In particular, the metric used is the $S$-index, defined as the flux contained within these two lines\footnote{When the Mt.~Wilson project started, large format digital detectors were not publicly available, so it was hard to get a digital spectrum, and it was also difficult to build a glass filter which such narrow bandwidth. Instead, a mechanical system was built to focus light from just the Calcium~H\&K lines and a region of the continuum. Such system used a rotating wheel placed in the focal plane of the spectrum and, as the wheel rotated, certain parts of the spectrum were allowed to pass and focused onto a single pixel detector. The triangular shape of the bandpass used in the definition of the $S$-index stems from the effect of the hole in the rotating wheel as it passed across the spectrum \citep{Vaughan1978}.} normalised by the nearby continuum \citep{Vaughan1978}. Another metric is the $\log R'_\mathrm{HK}$ index, which corrects the $S$-index for colour dependence and photospheric contribution, thus it is used to efficiently compare the activity level of stars with different spectral types \citep{Noyes1984}. For instance, the $\log R'_\mathrm{HK}$ index of the Sun at activity maximum and minimum is $-4.905$ and $-4.984$ \citep{Egeland2017}, while the same index for the M3.5 dwarf EV~Lac is $-3.75$ \citep{BoroSaikia2018}, demonstrating that the star is significantly more active than the Sun. The average level of these indices correlates with the activity level of the star, and their temporal variation is generally modulated at the stellar rotation period, therefore they can be efficiently used to constrain this fundamental parameter \citep[see e.g.][]{Baliunas1985b,Simpson2010,Hempelmann2016}.

There are also other activity indicators that rely on the amount of chromospheric emission in specific spectral lines, such as H$\alpha$, Na\textsc{I} D, and Ca\textsc{II} infrared triplet. The H$\alpha$ line is located at 6562.8\,{\AA} and its profile (for instance intensity and width) changes in response to processes within active regions on the star, such as plages, filaments, and flares. It is a well-studied activity indicator of magnetic activity in late-type stars, in contrast to solar-type stars, which are usually studied in the Ca~\textsc{ii} H\&K lines \citep{JoyAbt1974,Gizis2002,West2004}. The Na\textsc{I} D resonance lines are located at 5895.92\,{\AA} (D1 line) and 5889.95\,{\AA} (D2 line), and have been used complementarily to H$\alpha$. On one side the Na\textsc{I} D lines probe the conditions of the middle-to-lower chromosphere as opposed to H$\alpha$ that is sensitive to the upper chromosphere \citep{Mauas2000}, and on the other side, especially for M~dwarfs, the signal-to-noise ratio (S/N) in the spectral region of the Na\textsc{I} D lines spectral region is higher than for the Ca\textsc{II} H\&K lines. The Ca\textsc{II} infrared triplet is located at 8498.0\,{\AA}, 8542.0\,{\AA}, and 8662.0\,{\AA} and the corresponding activity indicator is defined based on the line central depressions \citep{Andretta2005}. These activity indices have been adopted interchangeably across different spectral types, assuming that they provide equivalent information on the stellar activity level. However, a number of studies have shown that the correlations between these indices are not ubiquitous and have an activity level dependence or a spectral type dependence, with M~dwarfs deviating from the general trends \citep[e.g.][]{GomesdaSilva2011,Martinez-Arnaiz2011,Meunier2024}. Finally, flaring activity is also generally used as an indicator of stellar activity \citep[see also Chapter 4 and the review by][]{Kowalski2024}.

The search for different activity indicators is dictated by the fact that their reliability depends on the spectral type of the star, for instance the classical Ca~\textsc{ii} H\&K is not optimal for M~dwarfs given their low flux emitted in the UV continuum, and spectroscopic exoplanet surveys have synergistically allowed the systematic monitoring of many spectral lines. Furthermore, several radial velocity exoplanet surveys are conducted in the near-infrared domain, hence the interest to define appropriate indicators. Compared to the optical domain, the near-infrared is a rather uncharted territory. Studies by \citet{Schofer2019} and \citet{Fuhrmeister2019} have shown that the He\textsc{I} triplet at 10830\,{\AA}, known to trace extreme ultraviolet irradiation from stellar coronae, is a useful indicator that correlates with H$\alpha$ emission and flaring events. \citet{Klein2021} demonstrated that it is a practical tracer of radial velocity variations at the stellar rotation period for the young, active M~dwarf AU\,Mic. In comparison, Pa$\beta$ traces different (more polar) regions, while Br$\gamma$ does not show rotational modulation. \citet{Terrien2022} investigated the K\textsc{I} line at 12435\,{\AA}, and reported rotationally-modulated variations consistent with photospheric magnetic fields changes. However, the K\textsc{I} line has neither sensitivity to flares, nor a correlation/anti-correlation with H$\alpha$ emission \citep{Fuhrmeister2022}. Other indicators based on specific spectral lines are also investigated \citep{CortesZuleta2023}. Finding new activity indicators is a complex task: different lines are sensitive to different active regions, and the diagnostic power of a line varies as the active region evolves with time. Therefore, the absence of a correlation with known activity tracers does not necessarily exclude the diagnostics power of a certain line. 

Moving outward from the chromosphere to the corona, stellar activity has been quantified at levels of the X-ray luminosity normalised by the bolometric luminosity of $L_X/L_\mathrm{bol}\sim10^{-8}-10^{-3}$ \citep{Vaiana1981,Schmitt2004,Gudel2004,Wright2010}. Young stars are the most X-ray luminous \citep{Telleschi2007}, while older stars have lower values reaching down to $L_X/L_\mathrm{bol}\sim10^{-8}$ \citep{Wright2010}. For young stars, accretion of material on the star can lead to a decrease in the X-ray emission due to the cooling of active regions \citep{Preibisch2005,Telleschi2007} and the circumstellar material can also absorb part of the emitted X-rays \citep{Flaccomio2003}. The decrease in X-ray emission with age can be attributed to the rotational spin-down over the lifetime of the star, which is driven by mass loss through a magnetized stellar wind \citep{Skumanich1972}.

Radio observations can trace the plasma conditions and magnetic field properties of stellar magnetospheres and coronae, thus providing powerful activity indicators \citep{Gudel2002,Yiu2024}. Radio and soft X-rays luminosities of magnetically active stars are known to correlate according to the G\"udel-Benz relation \citep{Gudel1993,Benz1994}, which spans ten orders of magnitude in radio luminosity. Physically, this means that there is a common mechanism underlying radio (non-thermal emission from accelerated electrons) and X‑rays (thermal emission from hot plasma) luminosities in magnetically active stars. For objects at the bottom of the main-sequence, the situation is different. In fact, observations of magnetic activity indicators in ultracool dwarfs revealed a significant decrease in X-rays emission \citep{Stelzer2006,Berger2010}, while the radio emission does not change \citep{Berger2001,Williams2014}. A possible explanation is found in the dichotomy of the large-scale magnetic field for late M~dwarfs (see Sect.~\ref{sec:results_activity}), for which stars with strong and axisymmetric magnetic fields obey the G\"udel-Benz relation while stars with weak, non-axisymmetric fields are radio over-luminous \citep{Williams2014}. 

\subsection{Magnetic fields}\label{sec:magfield_measurement}

Magnetic field measurements at the stellar photosphere rely mainly on the Zeeman effect \citep{Zeeman1897}, which occurs when atoms and molecules are subject to external magnetic fields. More details can be found in the lecture notes by \citet{Landstreet2009_1,Landstreet2009_2,Landstreet2009_3}. The quantum energy levels split in a certain number of sub-levels depending on the total angular momentum quantum number $J$, with associated spectral lines, in absorption or emission. Considering a normal Zeeman triplet the amount of splitting (in nm) is expressed by the following formula in CGS unit system
\begin{equation}\label{eq:Zeeman_split}
\Delta\lambda = \frac{e}{4\pi m_e c^2}\lambda_0^2g_\mathrm{eff}B = 4.67\cdot10^{-12}\lambda_0^2g_\mathrm{eff}B,
\end{equation}
with $e$ the elementary charge, $m_e$ the mass of the electron, $c$ the speed of light, $B$ the modulus of the magnetic field (in G), $\lambda_0$ the central wavelength in the absence of the field (in nm) and $g_\mathrm{eff}$ the Land\'e factor. The Land\'e factor is a dimensionless number spanning between 0 and 3 and represents the sensitivity to the magnetic field of a particular line. It can be computed using a simple analytical formula in the case of L-S coupling. Besides the Zeeman effect, there are other quantum mechanical effects resulting in energy changes of atomic levels that reflect in changes of line profiles and polarisation properties. Depending on the magnetic field strength, there are indeed different regimes: for atomic lines, between 50 and 100 kG we have the Paschen-Back regime (weakly magnetised degenerate stars), whereas above 100 kG (magnetised white dwarfs) we have the quadratic Zeeman effect. For non-degenerate stars on the main sequence with fields up to 10 kG these effects are not observable, although in some cases the Paschen-Back effect can be observed \citep{Crozet2023}.

\subsubsection{Magnetic field modulus and filling factor} \label{sec:Bmod}

Spectra in unpolarised light allow us to retrieve the total, unsigned stellar magnetic field. Assuming that the instrumental resolution is sufficient to distinguish the Zeeman-split components of individual spectral lines, and that the splitting is larger than the intrinsic line width, the idea is to measure the wavelength separation between the components and solve Eq.~\ref{eq:Zeeman_split} for $B$. 

When observing stars other than the Sun, further complications arise by the lack of spatial resolution, since observations convey only average information integrated over the stellar visible hemisphere, which is likely covered with inhomogeneities spanning a wide range of magnetic fields moduli and orientations. For unpolarised light, this means that we do not observe the Zeeman triplet clearly, but a more complicated and smeared pattern referred to as Zeeman broadening. This magnetic-induced broadening needs to be disentangled from non-magnetic line broadening (rotational, thermal, pressure, and instrumental). To disentangle Zeeman broadening, a possibility is to measure the width of individual atomic spectral lines formed under similar conditions, but with distinct magnetic sensitivities \citep{Saar1988, Reiners2012}. Otherwise, one can use different sets of diatomic molecules depending on the star's spectral type, especially for those characterised by forests of molecular lines that reduce the number of unblended atomic lines to only a handful \citep{Berdyugina2003,Afram2015}.

An additional phenomenon affecting magnetically sensitive, saturated or damping-regime spectral lines is called magnetic intensification, which leads to the desaturation of these strong lines \citep{Babcock1949,Basri1992}. In practice, the Zeeman-induced splitting of an atomic line reduces the optical depth at its centre and increases the optical depth at the wings. As a result, the line centre is desaturated and its equivalent width is larger. As studied by \citet{Stift2003}, magnetic intensification correlates with the magnetic field strength, and it has a complicated and strong dependence on the Zeeman splitting pattern, opacity, and line strength which cannot be expressed analytically. By comparing equivalent widths of multiple lines with different magnetic sensitivities, it is possible to measure the magnetic field modulus \citep{Basri1994,Kochukhov2020,Muirhead2020,Han2023}. In addition, \citet{Muirhead2020} recently showed the diagnostic power of the so called curve-of-growth diagrams, in which the equivalent width of lines belonging to the same multiplet is plotted against their absorption cross-section parametrised by $\log(gf)$ (with $f$ the oscillator strength and $g$ a statistical weight). Depending on the trends on the curve-of-growth diagram, one can identify lines that belong to the saturated or the damping regime, hence susceptible to Zeeman intensification.

\begin{figure}[!t]
    \centering
    \includegraphics[width=\textwidth]{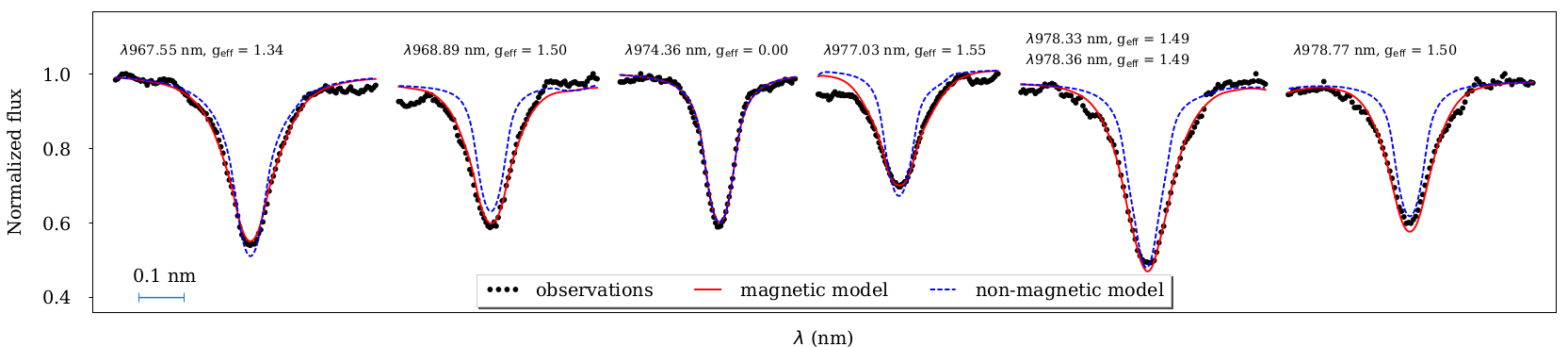}
    \caption{Theoretical modelling of Ti lines in the 965-680\,nm wavelength range, with different magnetic sensitivities for the M5V star V388~Cas. The star rotational velocity inferred from Ti lines is 11.7\,km\,s$^{-1}$ and the average magnetic field strength is 5.0\,kG. These lines are particularly useful because of the g$_\mathrm{eff}=0$ line, which is not magnetically broadened and therefore helps disentangling different broadening agents. A non-magnetic model (dashed blue line) and a magnetic multi-component one (red solid line) is overplotted to the observations (black dots). Image credit: \citet{Shulyak2019}.} 
    \label{fig:zbroadening}%
\end{figure}

For M~dwarfs in particular, spectrum synthesis modelling of the Ti I lines at 964.7-978.8 nm was adopted to perform extensive measurements (see Fig.~\ref{fig:zbroadening}), as they are excellent diagnostics for both Zeeman broadening and magnetic intensification \citep{Kochukhov2017,Shulyak2017,Shulyak2019}. The output of these analyses is encapsulated in the disk-averaged magnetic flux $Bf$ (measured in Gauss), where $B$ is the modulus of the field and $f$ is the filling factor in this context (not the oscillator strength), that is the portion of the stellar surface occupied by magnetic regions. Models postulate either a single value of $B$ and $f$ \citep{Saar1988} or a more elaborated multi-component, small-scale scenario \citep{Johns-Krull1999,Shulyak2014,Kochukhov2021}, but in general these two quantities are degenerate. More precisely, a strong magnetic field covering a small portion of the star can be equivalent to a weaker field covering a larger portion of the star. This degeneracy is reduced when observations of strong magnetic fields are conducted at high-resolution, in near-infrared and encompassing high Land\'e factor lines.

\subsubsection{Magnetic field vector} \label{sec:Bvec}

Owing to the Zeeman effect, circularly and linearly polarised spectra are sensitive to the magnetic field orientation, so they can be used to infer properties of the field geometry. In this case, the lack of spatial resolution translates into polarisation cancellation effects due to the contribution of areas with opposite field polarity, making the observations sensitive only to largest scales of the field. Therefore, repeated spectropolarimetric measurements are necessary to cover the rotation cycle of a star and eventually observe the magnetic field under different angles, namely a modulation of the polarity cancellation. This is a principal distinction with respect to Zeeman broadening analysis.

For active, cool main-sequence stars circular polarisation signatures are typically on the order of $10^{-3}$ the level of unpolarised continuum and linear polarisation signatures are one order of magnitude smaller \citep{Kochukhov2011,Lavail2018}. Observations of cool stars in linear polarisation mode are therefore limited to bright stars with strong surface magnetic fields. High-S/N observations are generally required to enable polarised Zeeman signatures to emerge from the noise and yield a reliable magnetic detection. In practice, for cool stars, this is achieved condensing the polarimetric information of the observed spectra into average line profiles using least-squares deconvolution \citep[LSD][]{Donati1997,Kochukhov2010a}. Owing to the often unknown magnetic sensitivity for molecular lines, atomic absorption lines are used almost exclusively for this purpose. The output average profiles benefit from a S/N gain that scales with the square root of the number of lines used in LSD and varies between a factor of 10 and 30 for M and G\,dwarfs. Modern echelle spectropolarimeters are effective at combining high resolving power with large wavelength coverage for this purpose.

\begin{figure}
    \includegraphics[width=0.85\textwidth]{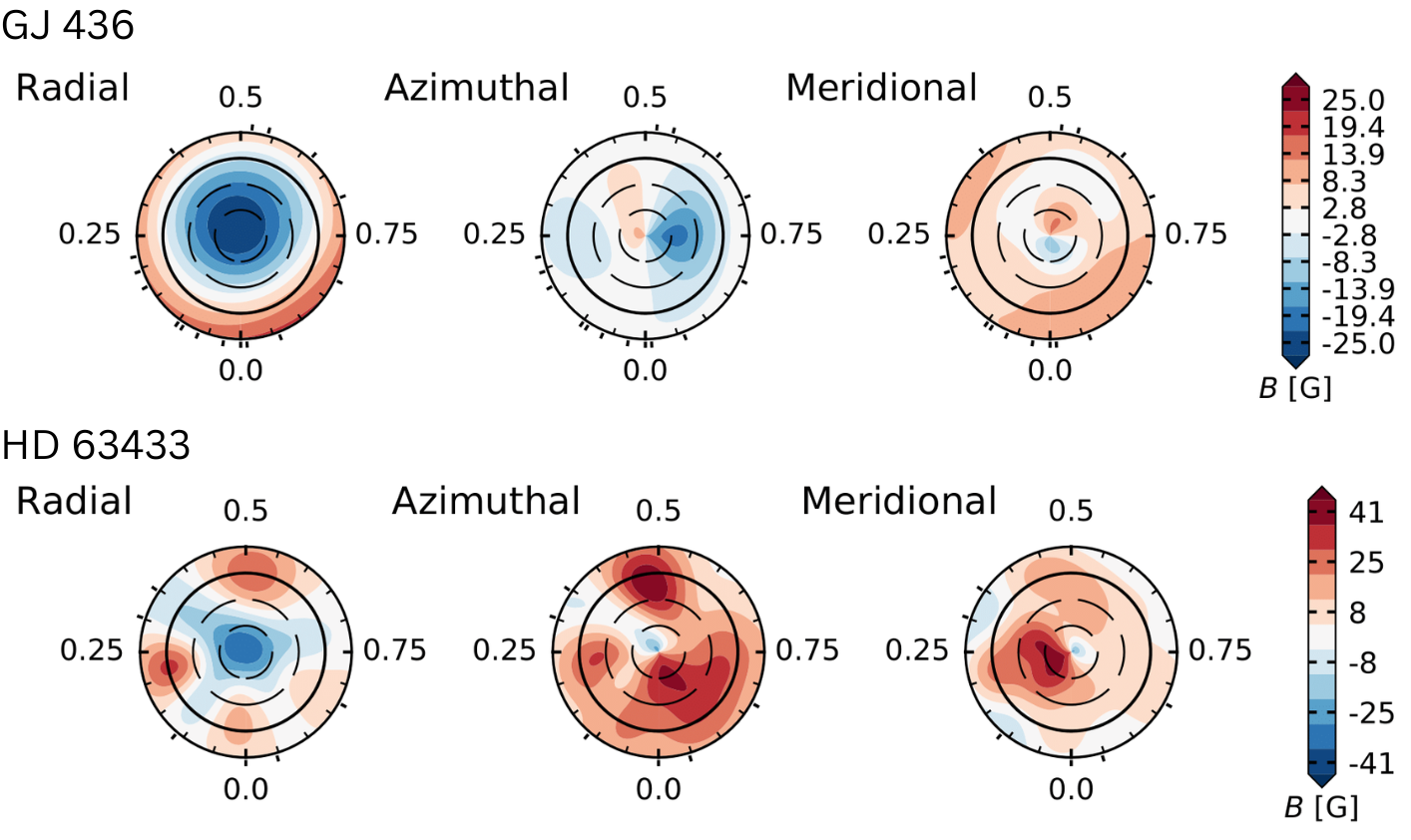}
    \centering
    \caption{Zeeman-Doppler imaging reconstruction in flattened polar view of the large-scale field of GJ\,436 and HD~63433. From the left, the radial, azimuthal, and meridional components of the magnetic field vector are displayed. The radial ticks are located at the rotational phases when the observations were collected, while the concentric circles represent different stellar latitudes: -30\,$^{\circ}$, +30\,$^{\circ}$, and +60\,$^{\circ}$ (dashed lines), as well as the equator (solid line). The colour bar indicates the magnetic polarity and the field strength (in G). GJ~436 (M2.5 type and 40.1~d rotation period) has a predominantly dipolar configuration, while HD~63433 (G5, 6.45\,d rotation period, and 400~Myr old) has a predominantly toroidal topology with a complex radial field, since the magnetic energy is distributed also in higher order spherical harmonics ($\ell=2$ and 3). Image credit: \citet{Bellotti2023b} and \citet{Bellotti2024b}.
        }\label{fig:zdi}
\end{figure}

To reconstruct the large-scale magnetic field topology of a star, a tomographic technique called Zeeman-Doppler imaging (ZDI; \citealt{Piskunov1983,Semel1989,DonatiBrown1997}) is used. The concept is similar to Doppler imaging (see Sect.~\ref{sec:starspots}), except that ZDI models rotationally modulated signatures of the magnetic field in circularly or linearly polarised light. The ZDI algorithm inverts a time series of polarised LSD profiles into a magnetic field map in an iterative fashion \citep[for more information see][]{DonatiBrown1997}. More precisely, synthetic line profiles are compared and updated with respect to the observed ones at each iteration, until convergence at a specific target $\chi^2_r$ is reached. Such a problem is ill-posed, meaning that infinite solutions could fit the observed data equally well, thus ZDI employs a regularisation scheme based on maximum entropy to choose a solution \citep{Skilling1984}. The algorithm searches for the maximum-entropy solution at a given $\chi^2$ level, that is, the magnetic field configuration compatible with the data and with the lowest information content. An example of reconstructed magnetic field topology for two exoplanet host stars is given in Fig.~\ref{fig:zdi}. 

The vector magnetic field is modelled as the sum of a poloidal and a toroidal component, both expressed via spherical harmonics decomposition \citep{Donati2006, Lehmann2022}, which is in contrast with earlier attempts where the field was reconstructed independently on each surface element. Such formal approach ensures that the reconstructed field is divergence-free, and it is effective at describing the properties of the large-scale magnetic field geometry (e.g., poloidal or toroidal, and axisymmetric or non-axisymmetric) and classify stars accordingly. The rotational velocity $v_\mathrm{eq}\sin(i)$ determines the maximum degree of the spherical harmonic expansion $l_\mathrm{max}$ in a proportional way, and therefore the amount of complexity we can image \citep{Hussain2009}. In other words, given a complex field topology, a higher value of $v_\mathrm{eq}\sin(i)$ will translate in a higher achievable $l_\mathrm{max}$ and therefore a more structured description of the field. Moreover, faster rotation and thus rotational broadening implies that the Zeeman signatures are more separated in radial velocity space, ultimately limiting polarity cancellation. Although the method initially relies on the fact the field is steady and observed variations are only attributable to rotational modulation, limited temporal variability can be accounted for in the form of solar-like differential rotation, expressed as
\begin{equation}\label{eq:diff_rot}
\Omega(\theta) = \Omega_\mathrm{eq} - d\Omega\sin^2(\theta),
\end{equation}
where $\theta$ is the colatitude, $\Omega_\mathrm{eq}=2\pi/\mathrm{P}_\mathrm{rot}$ is the rotational frequency at equator and $d\Omega$ is the differential rotation rate in rad\,d$^{-1}$. We note that differential rotation is also implemented in Doppler imaging, meaning that it is possible to constrain such parameter also from a time series of unpolarised light observations \citep[e.g.][]{Petit2002,Petit2004}.

\subsubsection{Caveats of magnetic field measurements}\label{sec:mag_limitations}

Both unpolarised spectroscopy and spectropolarimetry have provided numerous and fundamental results on the magnetic field for low-mass stars, but they are subject to limitations. In Table~\ref{tab:mag_limitations}, we summarise the properties of the magnetic field as retrieved by different techniques.

Zeeman broadening measurements output a value of the magnetic field strength regardless of its complexity, which provides partial information with which to feedback dynamo theories. Since the technique relies on precise radiative transfer modelling, its accuracy stems mostly from our knowledge of a line profile not affected by the magnetic field. The combined action of other line broadening agents needs to be carefully disentangled and prevents the measurement of weak fields, typically below a few hundred Gauss \citep{Anderson2010,Reiners2012}. For active (i.e. strong field), slowly rotating stars, magnetic broadening is more distinguishable from other broadening mechanisms and therefore its modelling is more favourable. As the rotational velocity increases, the limitation in measuring magnetic fields increases correspondingly.

Zeeman intensification measurements also output the magnetic field strength. Compared to Zeeman broadening, capturing the behaviour of line intensification requires the use of multiple lines with accurate transition probabilities and distinct responses to the magnetic field to distinguish the various agents affecting the lines strength. However, Zeeman intensification modelling techniques relies on equivalent width information, hence they have the advantage of alleviating the limitation at high stellar rotational velocity. For fast-rotating stars in fact, magnetic intensification measurements are used \citep{Shulyak2017,Muirhead2020,Cristofari2022a,Cristofari2022b,Han2023}.

Among the spectral types, M dwarfs have a greater number of observations because they are favourable targets to measure the total magnetic field: 1) the lines are overall narrower owing to a weaker temperature broadening, 2) their smaller size translates to a lower rotational broadening even at faster rotation rates, 3) the peak starlight emission is in near-infrared, giving access to a higher S/N for the measurements in combination to a larger broadening with respect to optical wavelengths, following Eq.\ref{eq:Zeeman_split}, and 4) even slower rotators (implying weaker rotational broadening) can host detectable and moderatively intense fields.

Zeeman-Doppler imaging outputs a map of the photospheric large-scale magnetic field. Polarity cancellation effects mask small-scale fields out, making local tangled features on the stellar surface undetected. In many cases, the brightness distribution or velocity fields (except for global rotation) are not accounted for, which are likely structured on small-scale like the magnetic field \citep{Rosen2012}. An advantage of polarisation spectra is that the null profile is simply a flat line at zero, hence the confusion with other broadening mechanisms does not appear. The application of tomographic techniques is less constrained by the stellar $v_\mathrm{eq}\sin(i)$ and measurements of weak fields on the order of a few G are feasible \citep{Auriere2009,Landstreet2015}. Very small $v_\mathrm{eq}\sin(i)$ values lead to more polarity cancellation \citep{Lehmann2024}, and although high $v_\mathrm{eq}\sin(i)$ values provide increased spatial resolution, there is a trade-off with detectability of signatures smeared by rotational broadening. Furthermore, even though retrieving the full information for spectropolarimetric data requires dense monitoring to apply ZDI, snapshot survey can also be successfully used to determine the presence of Zeeman signatures \citep{Grunhut2010,Auriere2013,Marsden2014,Auriere2015,Wade2016,Moutou2017}.

\begin{table*}[!t]
\caption{Summary of magnetic field measurement techniques.} 
\label{tab:mag_limitations}     
\centering                       
\begin{tabular}{l | l l l}      
\hline
{\bf Technique} & {\bf Zeeman broadening} & {\bf Zeeman intensification} & {\bf Zeeman-Doppler imaging} \\
\hline
Light type & Unpolarised & Unpolarised & Polarised \\ 
Line type & Single or multiplet & Multiplet & Average LSD profile \\
S/N & High & High & Ultra high \\
Spatial scale & Small and large & Small and large & Large\\
$\vec{B}$ property & Field strength & Field strength & Field topology \\
\hline                                
\end{tabular}
\end{table*}

A crucial aspect to consider is the complementarity of the two approaches: unpolarised light techniques convey information on the total magnetic field, while polarised light techniques give the orientation of the field on large-scales only. As a result, a significant contribution to the magnetic flux budget of a star is lost from polarimetry measurements \citep{Reiners2009,Kochukhov2019}. For M~dwarfs, as an example, the ratio of the field intensities derived from polarised and unpolarised light is lower than 20-25\% for fully-convective M~dwarfs, and lower than 10\% for partly convective ones \citep{Morin2010,Yadav2015}. This aspect, combined with the fact that the magnetic field strength measurement from Zeeman broadening modelling does not change across the two regimes, indicates that fully convective stars have magnetic regions with larger scales than partly convective ones. Capturing a complete view of stellar magnetic fields necessitates the usage of multiple techniques, hence an ideal scenario is when studies encompass both Zeeman broadening-intensification and Zeeman-Doppler imaging analyses.

\subsection{Observational results}\label{sec:results_activity}

\begin{figure}[t]
    \centering
    \includegraphics[width=\textwidth]{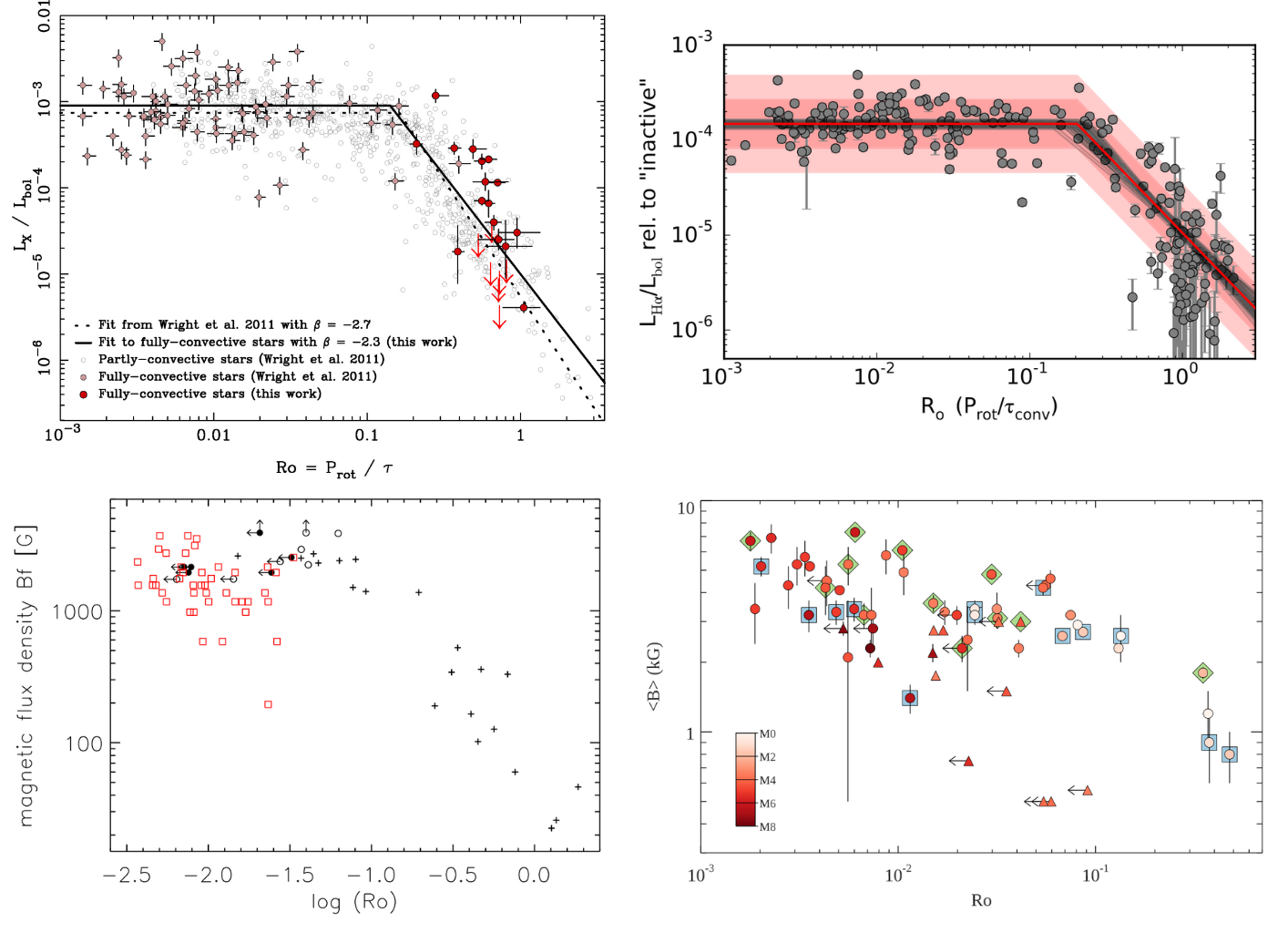}
    \caption[Rotation-activity relation seen from different activity proxies.]{Rotation-Activity relation for low-mass (later than mid-F) stars seen with different activity proxies. Top left: X-ray-to-bolometric index \citep{Wright2011}. Top right: H$\alpha$ index \citep{Newton2017}. Bottom left: Magnetic flux density \citep{Reiners2012}. Sun-like stars (crosses) M0-M6 (black circles), and M7-M9 (red squares) are shown. Bottom right: mean magnetic field obtained from radiative transfer modelling of Zeeman broadening \citep{Kochukhov2021}. Symbols are colour-coded by stellar spectral type, while the background markers indicate whether the magnetic topology is dipolar, axisymmetric (green diamonds) or multipolar, non-axisymmetric (blue squares). \label{fig:Xray_rotact}}
\end{figure}

The measurements of activity indicators and magnetic fields for several cool stars of F-M spectral types have shown interesting results and trends with stellar fundamental parameters, and in particular with the rotation period. Numerous studies have shown that these indicators follow the same activity-rotation relation \citep[e.g.][]{Noyes1984,Wright2011,McLean2012,Reiners2014,Newton2017,Wright2018}. This has been parametrised by the Rossby number ($Ro=P_\mathrm{rot}/\tau$): the ratio between the stellar rotation period and the convective turnover time. It is an empirical number, aimed at representing the fluid $Ro$ number which is a local quantity corresponding to the ratio between inertia and Coriolis forces. The latter is the typical timescale for a convective cell to rise in a fluid. As illustrated in Fig.~\ref{fig:Xray_rotact}, the activity indicators increase with decreasing Rossby number down to $Ro\simeq0.1$, symbolising a more efficient dynamo for faster rotators, whereas below $Ro\simeq0.1$ a saturated regime is reached.

Observations of main-sequence stars revealed distinct field intensities as inferred by Zeeman broadening and intensification modelling, on the order of 100 and 500\,G for Sun-like stars \citep{Basri1994,Valenti1995,Kochukhov2017,Hahlin2023}, and a factor of ten larger for active M dwarfs \citep{Linsky1985,Shulyak2019,Cristofari2022a,Cristofari2022b,Han2023}. There are also extremely intense cases such as WX\,UMa, with a field of 7.3\,kG \citep{Shulyak2017}. A dependence on the stellar rotation period similar to the activity indicators has been found for the magnetic field strength \citep{Saar1996,Shulyak2017,Kochukhov2021} which, in turn, exhibited a secondary dependence on magnetic field large-scale field geometry \citep{Shulyak2017,Shulyak2019}. Stars with multipolar fields saturate around $P_\mathrm{rot}$=4\,d and $B$=4\,kG, while stars with dipolar fields can sustain stronger fields at faster rotation (see Fig.~\ref{fig:Xray_rotact}). Given the large scatter of the data points at short rotation periods, concluding that a maximum magnetic field is achieved or that the field could still increase is not possible, which motivates more observations of ultra-rapid rotators. The temporal variability of these indicators may also be significant enough to increase scatter and blur such relation. From field strength measurements overall, it seems that a change in the relation exists for late M~dwarfs. 

In terms of internal structure, there is no abrupt change in X-ray emission or magnetic field strength at the transition between partly- and fully-convective \citep{Reiners2012,Wright2018}, suggesting that at least the energetics of the underlying dynamo mechanisms may be similar, following the scaling law reported in \citet{Christensen2009}. Instead, we observe a possible breakdown of the activity-rotation relation around spectral type M7-M9 since, given a $Ro$ value, the distribution of field strength for these stars reaches lower values than the earlier counterparts \citep{Reiners2009}.

As far as spectropolarimetric observations are concerned, Zeeman-Doppler imaging has revealed a diversity of magnetic field topologies and properties. Partly-convective, fast-rotating stars with mass above 0.5\,M$_\odot$ tend to have moderate, predominantly toroidal and non-axisymmetric large-scale fields, whereas between 0.35 and 0.5\,M$_\odot$ they are stronger, poloidal and axisymmetric \citep{Donati2008,Morin2008,Phan-Bao2009}. For stars close to the fully-convective limit, the field topology is dipolar. For fully-convective stars below approximately $0.2$\,M$\odot$, a dichotomy of topologies has been observed to co-exist: weak, complex, non-axisymmetric against strong, simple, axisymmetric \citep{Morin2010}. An evident example of this dichotomy is GJ\,65, a binary system consisting of two M~dwarfs (M5.5 and M6) with nearly identical masses, rotation period and evolutionary track, but with two distinct topologies: weak, non-axisymmetric (GJ\,65\,A) and strong, axisymmetric (GJ\,65\,B) \citep{Kochukhov2017}. Such results, together with the scatter of field strengths for stars with similar $Ro$, symbolise that dynamo efficiency is not solely determined by stellar rotation period and mass, as predicted in the \citet{Morin2011} strong field dynamo idea, but an unknown combination of fundamental parameters is at play \citep{Kochukhov2019}.

The findings on fully-convective stars can be explained by either two distinct and independent branches of dynamo (known as bistability, \citealt{Morin2012,Gastine2013}) or by assuming the presence of magnetic cycles for which we reconstruct only a snapshot of a long-term topological variation \citep{Kitchatinov2014}. This motivates long-term spectropolarimetric monitoring of active stars further, in order to trace any temporal evolution of the large-scale field geometry and shed more light on which of the two hypotheses is more plausible. In addition, there is increasing interest in making magnetic field maps for slower rotators, both Sun-like \citep{Petit2008,Metcalfe2021,Metcalfe2024} and M~dwarfs \citep{Hebrard2016,Moutou2017,Bellotti2023b,Lehmann2024}. On one side, this will cover an uncharted parameter space and enlarge the parameter space to feed back dynamo theories with, and on the other side it will give access to crucial information to understand the stellar environment in which exoplanets are embedded \citep[e.g.][]{Vidotto2015,Garraffo2016,Vidotto2021,Alvarado-Gomez2022b}.

\subsection{Stellar dynamo models}\label{sec:stellar_dynamo}

Between mid-F and M3 spectral type, stars are partly-convective and share similar internal structures as the Sun (with varying aspect ratios). For spectral types later (colder) than M3, stars are fully-convective and have no tachocline. In terms of stellar mass, the theoretical limit between these two regimes is 0.35\,M$_\odot$ \citep{Chabrier1997} and it is in agreement with observations, since it was used to explain the {\it Gaia} magnitude gap \citep{Jao2018,Feiden2021}. However, metallicity affects the depth of the convective envelope \citep{VanSaders2012,Tanner2013}, and strong magnetic fields may quench convection, pushing the boundary towards lower masses \citep{Mullan2001}. For this reason, the value of 0.35\,M$_\odot$ should not be considered as a sharp boundary. Given the distinct interior structures, extending dynamo studies to other stars is fundamental to contextualise the solar dynamo and understand how the generation of magnetic fields vary over different stellar temperatures, convective velocities, rotation rates, and Rossby number.

Generally, a solar-like $\alpha\Omega$ dynamo is adopted to explain the generation of magnetic fields in partly-convective stars. Quantifying the influence of key parameters such as stellar rotation and mass in characterizing the dynamo and magnetic level achieved in solar-like stars is not straightforward, and requires multi-dimensional numerical simulations to account for the excitation of different types of convective dynamos that may occur \citep[see e.g.][]{Brun2017,Charbonneau2020,Kapyla2023}. Some studies have applied the solar mean-field dynamo paradigm to other spectral types, while others performed global 3D MHD simulations to model differential rotation and stellar magnetism of solar-like stars. Results indicated the emergence of magnetic cycles in some parameter regimes \citep[see][for example]{Guerrero2016,Strugarek2018,Viviani2019}. Simulations by \citet{Brun2022} of solar-like convective dynamos with distinct masses and rotation periods showed long magnetic cycles for small fluid Rossby numbers, while other studies obtained irregular patterns for fast rotators \citep{Vashishth2023}. Recent numerical simulations by \citet{Noraz2024} showed that fast rotators tend to exhibit rapid evolution and local polarity reversal. 

Numerical MHD simulations for early-type, rapidly-rotating M~dwarfs performed by \citet{Bice2020} indicated that a tachocline is not necessary to achieve strong and organised toroidal wreaths, although its presence would regularise magnetic cycles and allow long-term variations. Turning to fully convective stars, early ideas on dynamo action indicated that the $\Omega$-effect cannot apply since the stars exhibit solid body rotation \citep{Kuker1999}, and do not possess a tachocline, where the $\Omega$-effect is supposed to be the most efficient. An early idea for the generation of magnetic fields required invoking small-scale dynamo, that is at the scale of plasma motions \citep{Durney1993}, while scenarios in which cyclonic turbulence is the sole mechanism at play have been studied recently (i.e. $\alpha^2$ models, \citealt{Chabrier2006,Yadav2015}). Simulations in this case yield strong (on the order of kG) large-scale magnetic fields with a significant axisymmetric component \citep{Dobler2006,Browning2008}, but fail at reproducing the observations (see Sect.~\ref{sec:cycles}). There has been extensive effort to understand the generation of large-scale fields dominated by an axisymmetric dipole, especially for large density contrasts \citep{Gastine2012,Raynaud2015,Yadav2015,Zaire2022}. In particular, we also note that the work of \citet{Yadav2015} reconciles the observations of small- and large-scale magnetic fields. Alternatively, mean-field dynamo models are capable of predicting long-term evolution of the field geometry, but provide no information on the global field strength \citep{Kitchatinov2014,Pipin2017}. \citet{Yadav2016} carried out MHD simulations tailored to be representative of Proxima Cen and predicted a 9-yr cycle of both field strength and topological variations, in accordance with the 8-yr cycle seen with photometry \citep{SuarezMascareno2016,Wargelin2017,Wargelin2024}.

\section{Stellar cycles}\label{sec:cycles}

Observing magnetic cycles in other stars than the Sun provides key constraints to dynamo theories, and in particular how fundamental stellar parameters, such as mass and rotation period, impact the internal dynamo processes \citep[][for recent reviews]{Jeffers2023,Charbonneau2023}. In the following, we summarise the observational techniques used to investigate the presence of stellar cycles and the main trends.

\subsection{Observational techniques}\label{sec:techniques}

Long-term photometric time series can reveal stellar Schwabe cycles associated to the evolving distribution and surface coverage of surface inhomogeneities like spots and faculae \citep{Baliunas1985,Strassmeier2009}. Observations have been conducted from both ground- and space-based facilities, reaching baselines on the order of 30-40\,yr. The analyses showed a variety of long-term modulations of the light curves, from regular, to multiple, to irregular \citep{Olah2000,Olah2009,Messina2002,Savanov2012,FerreiraLopes2015,Lehtinen2016}. \citet{SuarezMascareno2016} investigated the light curves of 47 stars and found cyclic periodicities between 2 and 14\,yr. Dividing their sample by spectral type (from F to mid-M), they showed that the distributions of cycle period did not exhibit significant differences. In addition, they showed a clear decrease in photometric amplitude induced by the cycle with increasing stellar rotation period. Making a distinction based on age for a sample of G-K dwarfs. \citet{Olah2016} noted that old, slowly-rotating stars tend to have smooth and simple cycles, whereas younger fast-rotating stars exhibit abrupt changes and complex temporal variations. The age division between stars with smooth and complex cycles is at around 2 to 3\,Gyr. 

\citet{Vida2014} and \citet{Reinhold2017} analysed $Kepler$ light curves spanning four years for stars whose rotation period is $<1$~\,d and between 1-40\,d, respectively, and found activity modulations with time scales between 0.5 and 6\,yr. The measurements indicated a correlation between the cycle period and the rotation period. Some short-term variations may be due to random fluctuations owing to degeneracies associated to starspots distributions \citep{Walkowicz2013,BasriShah2020}, lifetimes and differential rotation, therefore putting some of the short cycles into question. For M dwarfs, the analysis of photometric light curves by \citet{Savanov2012} did not reveal a dependence of activity cycles on the rotation period, which was also reported later by \citet{Irving2023} for fully-convective M dwarfs. It is important to note that the temporal baseline with which stellar cycles on M dwarfs have been investigated is shorter than for the other spectral types, an aspect that can influence the correlation analysis between cycle period and rotation period.

Asteroseismic observations can also reveal the presence of stellar magnetic cycles. Indeed, p-mode (acoustic) oscillations propagating in magnetic regions are sensitive to the temporal changes of the physical conditions in such regions, as shown for the Sun \citep[e.g.][]{WoodardNoyes1985,Basu2016,Broomhall2017,Betrisey2024}. The first detection of activity-related p-mode variations on another star was reported by \citet{Garcia2010} for the F dwarf HD~49933 using $CoRoT$ data. The idea is to look for the imprint of the magnetic cycle on parameters of p-mode oscillations, namely the frequency and amplitude, in the form of temporally correlated or anti-correlated modulations with other activity diagnostics \citep{Mathur2013,Salabert2016,Kiefer2017,Karoff2018}. The work of \citet{Santos2018} on a large sample of $Kepler$ solar-like stars (T$_\mathrm{eff}=5000-6600$\,K) revealed correlations between the magnetically-induced shifts in acoustic frequency with fundamental stellar parameters such as rotation period, age, and activity (that is $S$-index, see Sect.\ref{sec:measure_activity}).

Another approach is to monitor the fluctuation of stellar activity over several years. At the chromospheric level, one can track the variation of the heating in the stellar atmosphere as conveyed by the emission reversal in the cores of chromospheric lines \citep{Leighton1959,Hall2008}, like the Ca\textsc{ii} H\&K lines described in Sect.~\ref{sec:measure_activity}. Using the $S$-index, it was possible to monitor the long-term variations of magnetic activity of hundreds of FGK stars, finding the presence of activity cycles \citep{Baliunas1995}. Regular cycles with time scales shorter (5.7\,yr, HD~161329) or longer (21\,yr, HD~219834\,A) than the solar sunspot cycle were reported, together with cases exhibiting the superposition of multiple cycles, such as HD~76151 and HD~190406 having $\sim$2.5\,yr variations modulated over 16-17\,yr. Irregular variations (e.g. HD~61421), flat time series (e.g. HD\,124570) were reported, together with Maunder-like minima from more recent works \citep[HD~4915 and HD~166620 by][]{Shah2018,Baum2022,Luhn2022}. These results are consistent with the photometric diagnostics. Monitoring activity cycles for solar-like stars with different ages can inform about the temporal variability of the younger Sun, such as the case of the solar twin 18~Sco \citet{doNascimento2023}. The star is younger than the Sun (3.4-3.7\,Gyr) and exhibits a faster magnetic cycle of 15~yr (see also Sect.~\ref{sec:solar_activity}).

\begin{figure}[t]
    \centering
    \includegraphics[width=\textwidth]{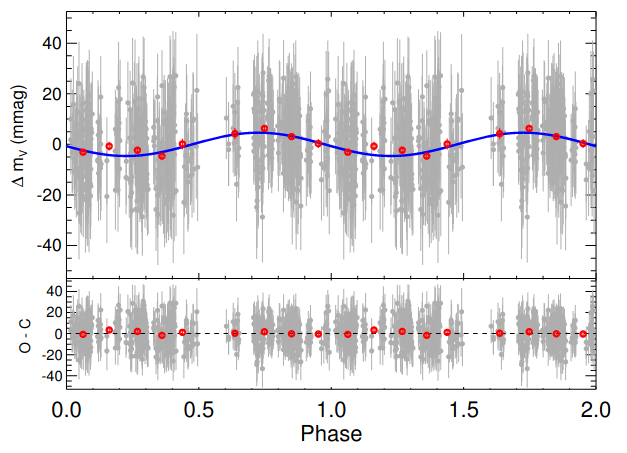}
    \caption{Long-term photometric cycle of HD~2071, phase-folded at a cycle period of $11.2\pm3.6$~yr with a semi-amplitude of $4.9\pm1.3$~mmag. Grey dots show individual measurements, while red dots show the average measurement of every phase bin. Image credit: \citet{SuarezMascareno2016}. \label{fig:phot_cycle}}
\end{figure}

Studies of S-index variations extended the temporal baseline to investigate cycles \citep{Olspert2018,Baum2022,Mittag2023} and included M dwarfs as well \citep{BoroSaikia2018,Isaacson2024}, confirming the variety of long-term variability. In terms of stellar fundamental parameters, \citet{Isaacson2024} provided tentative evidence for a range of stellar activity in which nearly every G and K type manifests cycles: effective temperature in the range 4700-5900~K and $\log R'_\mathrm{HK}$ between -4.9 and -4.7. In addition, they showed that the amplitude of the cycle is larger for later spectral types and younger stars, with M dwarfs having the largest amplitude without a clear dependence on age. For M~dwarfs specifically, other chromospheric indicators have also been used (and sometimes compared) to look for cycles, in particular H$\alpha$. Monitoring has often been carried out in the framework of radial velocity planet-search programs like \citet{GomesDaSilva2012,Robertson2013,Mignon2023} or for their preparation such as \citet{Isaacson2024}.

In a similar way to chromospheric heating, stellar cycles can be identified by the variability of coronal heating, as observed in X-ray emission \citep[][]{Gudel2004}. However, only a limited number of stars have been monitored in X-rays, primarily due to the challenge in performing decades-long campaigns with space-based telescopes operating at these wavelengths. The first star observed to have a coronal activity cycle is 61\,Cyg~A \citep{Hempelmann2006,Robrade2012}, which also reflected the chromospheric activity cycle present on the star. Additional examples are $\alpha$~Cen~A \citep[$\sim12-15$\,yr cycle][]{DeWarf2010}, HD~81809 \citep[7.3\,yr cycle][]{Favata2008,Orlando2017}, $\alpha$~Cen~B \citep[8.84\,yr cycle][]{DeWarf2010,Robrade2012}, $\iota$~Hor \citep[1.6\,yr cycle][]{SanzForcada2013}, and $\varepsilon$~Eri \citep[2.9\,yr cycle][]{Coffaro2020}. The spectral types of these stars range from to F8 to K5, and have known chromospheric cycles that feature a correlated variation.

Two additional methods to look for activity cycles rely on radio emission and flare statistics. In the first case, \citet{Route2016} reported reversals of polarised radio emission for ultra-cool dwarfs, while \citet{Bloot2024} showed evolution of polarised radio emission of AU~Mic, possibly suggesting an underlying magnetic cycle. In the second case, recent studies have claimed the potential of using variations in flare statistics as probes for stellar cycles \citep{Feinstein2024,Wainer2024,Raetz2024}, which is reminiscent of phenomena observed on the Sun \citep{Rieger1984,Veronig2002,Bai2003}.

Searching for magnetic cycles can be performed with dedicated long-term spectropolarimetric campaigns, aiming to capture the evolution of the large-scale magnetic field topology. Campaigns like BCool \citep{Marsden2014} have now reached baselines of 15-20\,yr, which is suitable for inspecting the yearly evolution of the magnetic topology with ZDI. Different studies have reported the existence of magnetic cycles across several spectral types. Among F-type stars, $\tau$~Boo has likely the densest monitoring \citep{Catala2007,Donati2008,Fares2009,Fares2013}. Initial spectropolarimetric observations revealed a 2-yr magnetic cycle with clear polarity reversals, while subsequent observations by \citet{Jeffers2018} constrained the time scale to be 120\,d, which is in agreement with $S$-index variations \citep{Mengel2016}. \citet{Vidotto2012} showed that the absence of a detectable X-ray cycle is consistent with spectropolarimetric results as well. Other F-type stars showing polarity reversals are HD78366 \citep{Morgenthaler2011} and HD~75332 \citep{Brown2021}, while $\chi$~Dra did not show significant signs of evolution of 5\,yr \citep{Marsden2023}. 

For G-type stars, $\kappa$~Cet exhibits a Hale-like cycle over a time scale of 10\,yr, characterised by polarity reversals in both the radial and azimuthal field, with oscillations in topology complexity and axisymmetry \citep{doNascimento2016,BoroSaikia2022}. Other interesting cases are HD\,19077, for which a polarity reversal in azimuthal field occurred, followed by a reversal of the radial field only in subsequent epochs \citep{Petit2009,Morgenthaler2011}, and V889~Her, that showed only attempts at polarity reversals \citep{Brown2024}. \citet{Bellotti2025} examined six solar-like stars and found: repeating polarity reversals in the radial or toroidal field component for HD~9986 and HD~56124, one polarity reversal in the toroidal component for HD~73350, a short-term evolution (2.5 yr) modulated by the long-term (16 yr) variation for HD~76151 consistently with the reported chromospheric cycle, and no evident cyclic evolution for the young stars HD~166435 and HD~175726. For other G-type stars with spectropolarimetric observations like HN~Peg, EK~Dra, and HD~171488, cyclic variability has also not been captured so far \citep{BoroSaikia2015,Waite2017,Marsden2006,JeffersDonati2008,Jeffers2011,Willamo2022}

For K-type stars, 61~Cyg~A manifests a magnetic cycle over 7.3\,yr, with polarity reversals of the radial and azimuthal field, in phase with the chromospheric and X-ray cycle \citep{BoroSaikia2016,BoroSaikia2018b}. There is also the case of $\varepsilon$~Eri, with a complex temporal evolution of the topology \citep{Metcalfe2013,Jeffers2014,Petit2021,Jeffers2022}, likely reflecting the superposition of the two known cycles from $S$-index monitoring, and LQ~Hya, with one polarity reversal detected potentially coincident with the $S$-index minimum \citep{Lehtinen2022}.

\begin{figure}[t]
    \centering
    \setlength{\tabcolsep}{2pt}
    \captionsetup[subfigure]{position=top,singlelinecheck=off} 
    \captionsetup[subfigure]{labelformat=empty}
    \begin{tabular}{cccccc}    
    \subfloat[2007.59]{\includegraphics[width=0.16\textwidth]{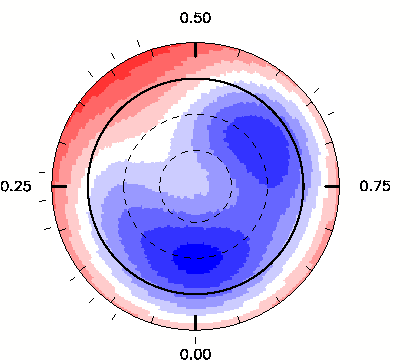}}&
    \subfloat[2008.64]{\includegraphics[width=0.16\textwidth]{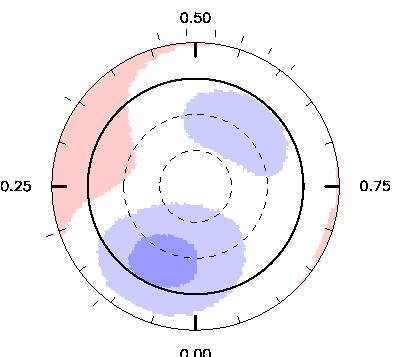}}&
    \subfloat[2010.55]{\includegraphics[width=0.16\textwidth]{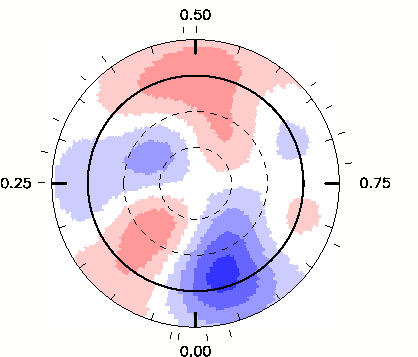}}&
    \subfloat[2013.61]{\includegraphics[width=0.16\textwidth]{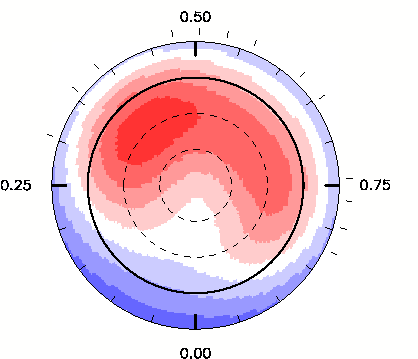}}&
    \subfloat[2014.61]{\includegraphics[width=0.16\textwidth]{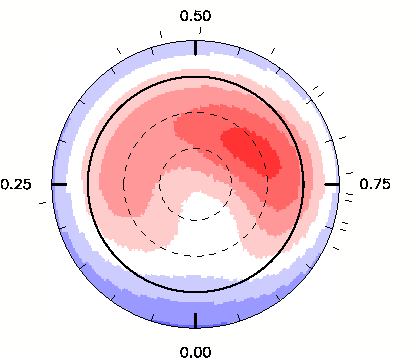}}&
    \subfloat[2015.54]{\includegraphics[width=0.16\textwidth]{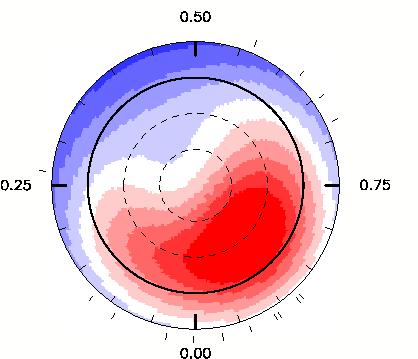}}\\
    \subfloat[]{\includegraphics[width=0.16\textwidth]{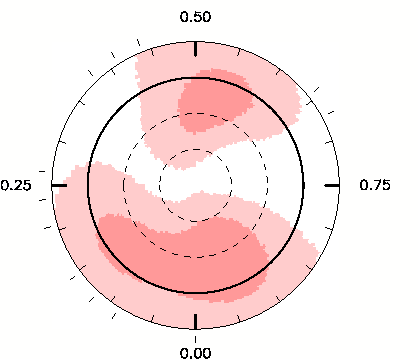}}&
    \subfloat[]{\includegraphics[width=0.16\textwidth]{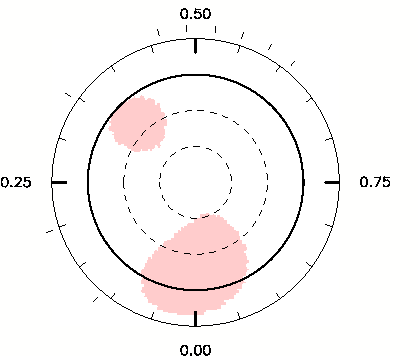}}&
    \subfloat[]{\includegraphics[width=0.16\textwidth]{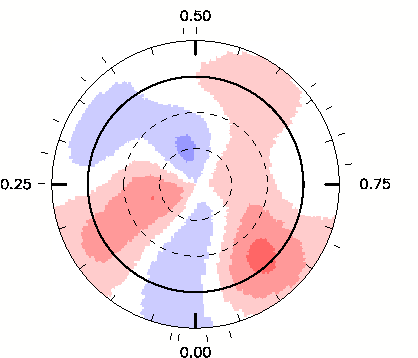}}&
    \subfloat[]{\includegraphics[width=0.16\textwidth]{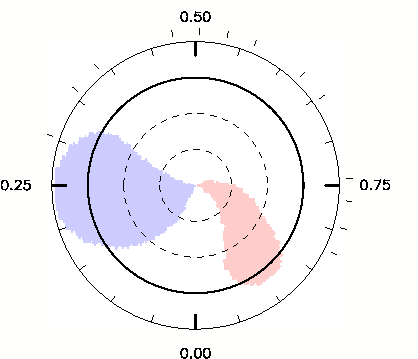}}&
    \subfloat[]{\includegraphics[width=0.16\textwidth]{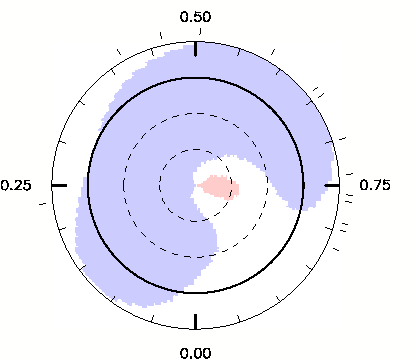}}&
    \subfloat[]{\includegraphics[width=0.16\textwidth]{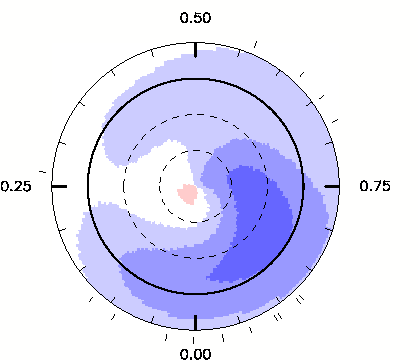}}\\
    \end{tabular}
    \includegraphics[scale=0.25, angle=270]{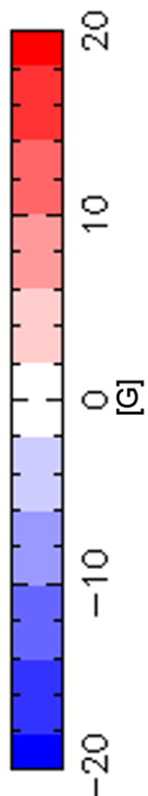}
    \caption{Reconstructed large-scale magnetic field of 61~Cyg~A across six different epochs. The first row indicates the radial component of the magnetic field and the second row illustrates the azimuthal field. The format of the panels is the same as Fig.~\ref{fig:zdi}. Evidence of a polarity reversal of the radial field can be seen between epoch 2007.59 and 2013.61. Image credit: \citet{BoroSaikia2016}. \label{fig:61cyg_cycle}}
\end{figure}

For M-type stars, spectropolarimetry studies have not captured a full magnetic cycle, but there is evidence that these stars undergo such phenomena. A clear example is GJ~1151, for which \citet{Lehmann2024} reported a first polarity reversal, and others showed significant variations in axisymmetry and magnetic field strength such as GJ~388, GJ~873, GJ~905, GJ~1289, \citep{Bellotti2023,Lehmann2024,Bellotti2024}. Another case is GJ~410, that switched from a predominantly toroidal to a poloidal configuration \citep{Donati2006,Hebrard2016,Bellotti2024}. There are also cases for which no signifcant evolution is observed besides an oscillation in field strength, such as GJ~406, GJ~408, GJ~617~B, and GJ~1286 \citep{Bellotti2024,Lehmann2024}.

In the context of exoplanets, evidence of activity cycles was also found via radial velocity exoplanet searches \citep{GomesDaSilva2012,LopezSantiago2020}. Studies established that cycles introduce long-term signals in radial velocity data sets that can dominate over planetary signatures, given that they modulate the appearance and number of heterogeneities on the stellar surface. The dominant mechanism is the magnetic inhibition of convective blueshift, that evolves along the cycle \citep{Meunier2010, Meunier2019}. It is therefore necessary to have an accurate constraint on the temporal variations of the cycle, in order to remove its signature and allow a more reliable planetary detection and characterisation \citep{Lovis2011,Costes2021,Sairam2022}. Reciprocally, radial velocity long-term evolution becomes an additional activity proxy that one can use to track activity cycles.

\subsection{General trends}\label{sec:cycle_trends}

Searching for stellar activity cycles benefits from a multi-wavelength approach, since the methods described in Sect.\ref{sec:techniques} probe the effects of dynamo action across different layers of the stellar atmospheres and interiors. Photometric monitoring has been one of the most prolific methods, capturing on the order of 3300 short- and long-term cycles\citep{Vida2014,Ferreira-Lopes2015,SuarezMascareno2016,Lehtinen2016,Clements2017,Reinhold2017}. This is followed by the chromospheric monitoring via $S$-index diagnostics, scoring between 300-400 cycle detections \citep{Baliunas1995,BoroSaikia2018,Baum2022,Isaacson2024} and 100-200 by asteroseismic analysis \citep{Garcia2010,Mathur2013,Regulo2016,Salabert2016,Kiefer2017,Santos2018,Betrisey2024}. From X-rays and spectropolarimetric campaigns, around a dozen detections of cycles were reported. We note that spectropolarimetry is sensitive to magnetic (Hale) cycles, while most of the other techniques are sensitive to activity (Schwabe) cycles.

On the Sun, the various activity proxies used by these methods have shown correlated temporal modulations over the magnetic cycle. For instance, the brightness variations are correlated to the evolution of chromospheric emission lines \citet{White1981}, the $S$-index \citep{Radick2018}, and the soft X-rays modulation \citep{Acton1996,Judge2003,Ayres2020}. Variations in acoustic ($p$-mode) frequencies and amplitudes follow the magnetic cycle in a correlated and anti-correlated manner, respectively \citep{Chaplin2001,Broomhall2017,Kiefer2018}. Finally, as seen already in Fig.~\ref{fig:sun_field}, the large-scale magnetic field topology of the Sun is simple and dipolar at $S$-index maximum and more complex, that is with more energy in the quadrupolar mode at activity maximum \citep{Vidotto2018,Lehmann2021}.

For other stars, \citet{Radick1998} analysed contemporaneous photometric and chromospheric Ca\textsc{ii} H\&K emission time series measurements, finding that younger, more active stars become fainter with increasing Ca\textsc{ii} H\&K emission, while older, less active stars become brighter with increasing Ca\textsc{ii} H\&K emission, which is what the Sun does during the activity cycle. These trends were confirmed also by a subsequent study with data sets spanning a larger number of years \citep{Radick2018}, and can be interpreted in terms of spot-dominated activity in young stars against faculae-dominated activity in older stars. Cases in which the photometric variations are synchronised with the chromospheric emission have also been found, like HD~30495 \citep{Soon2019}. There are also stars whose large-scale magnetic field evolution is correlated with the $S$-index, with a simple dipolar topology at activity minimum and a complex one at activity maximum 61~Cyg~A \citep{BoroSaikia2016,BoroSaikia2018b}. However, the link between $S$-index and large-scale diagnostics is not always straightforward, because activity indices are computed from unpolarised spectra, and are therefore sensitive to most magnetic features both on small and large spatial scales. For this reason, moderate correlations or more complex behaviours between these quantities have been reported for cool dwarfs \citep[e.g.][]{Marsden2014,Brown2022}. Recently, \citet{Wargelin2024} showed that the UV and the X-ray intensities of Proxima~Cen are anti-correlated with optical brightness modulations, in stark contrast to the Sun. 

Over time, additional trends have been reported from stellar cycle searches. Young, fast-rotating stars tend to have irregular or chaotic variations, with a fast evolution, while older, slowly-rotating stars exhibit smoother cycles with less pronounced amplitudes in their variability \citep[see e.g.][]{BoroSaikia2018}. There are also distinct cases, showing the superposition of multiple cycles with different time scales. In terms of cycle shape, that is the ascending and descending phase lengths, \citet{Willamo2020} found similarities with respect to the Sun, albeit the solar cycle represents a particularly asymmetric case. 

With the increasing number of detected stellar cycles, a relationship between rotation and cycle periods started to emerge \citep{Noyes1984b}. It shows different scaling branches, named `active' and `inactive', with the active stars showing longer cycles than inactive ones at a fixed rotation period \citep{Brandenburg1998,saar1999,Lehtinen2016}. These branches may be symbolic of different dynamo mechanisms operating either near the surface or at the bottom of the convection zone \citep{Bohm-Vitense2007}. The Sun falls in-between the branches, potentially indicating a transition dynamo state between them \citep{Metcalfe2016}. The idea is that activity cycles initially increase in length as the stellar rotation period slows over time, but when stars reach a critical Rossby number, weakened magnetic braking sets in and the rotation period remains nearly constant while the cycle grows longer and weaker before disappearing entirely \citep{Metcalfe2017,Metcalfe2022}. However, additional observations have put these interpretation into question \citep{Savanov2012,See2016,Strugarek2017,Reinhold2017,Olspert2018,BoroSaikia2018,Amard2020,Bonanno2022,Mittag2023}, therefore the existence of these branches is debated.\\\\

{\it Acknowledgements:} SB acknowledges funding by the Dutch Research Council (NWO) under the project ``Exo-space weather and contemporaneous signatures of star-planet interactions" (number OCENW.M.22.215 of the research programme ``Open Competition Domain Science- M").

% Alphabetically ordered of the bibliography.
%\bibliographystyle{aa}
\bibliographystyle{apalike}
\bibliography{bibliography}
%\printbibliography %line used if we're using biblatex and not natbib

\end{document}